\documentclass[aps,preprint,english]{revtex4}
\usepackage[T1]{fontenc}
\usepackage[latin1]{inputenc}
\usepackage{float}
\usepackage{graphicx}
\usepackage{pxfonts}
\usepackage{verbatim} 

\makeatletter

\usepackage{graphicx}

\usepackage{babel}
\usepackage{setspace}

\makeatother
\begin{document}
\title{Transition-Event Durations in One Dimensional Activated Processes}
\author{Bin W. Zhang and David Jasnow \footnote{Electronic mail: jasnow@pitt.edu}}
\affiliation{Department of Physics \& Astronomy, University of Pittsburgh, Pittsburgh, Pennsylvania 15260}
\author{Daniel M. Zuckerman \footnote{Electronic mail: dmz@ccbb.pitt.edu}}
\affiliation{Department of Computational Biology, School of Medicine, University of Pittsburgh, Pennsylvania 15213}
\date{\today}
\begin{abstract}
Despite their importance in activated processes, transition-event
durations --- which are much shorter than first passage times ---
have not received a complete theoretical treatment.  We therefore study
the distribution $\rho_{b}(t)$ of durations of transition events over
a barrier in a one-dimensional system undergoing over-damped
Langevin dynamics.  We show that $\rho_{b}(t)$ is determined by a
Fokker-Planck equation with absorbing boundary conditions, and
obtain a number of results, including:  (i) the analytic form of
the asymptotic short-time behavior ($t \rightarrow 0$), which is
universal and independent of the potential function;
(ii) the first non-universal correction to the
short-time behavior; (iii) following Gardiner
~\cite{Gardiner-Stochastic_Methods_Handbook-E2nd-1994},
a recursive formulation 
for calculating, exactly, all moments of $\rho_{b}$ based solely on the potential function
--- along with approximations for the distribution based on a small
number of moments; and (iv) a high-barrier approximation to the
long-time ($t \rightarrow \infty$) behavior of $\rho_{b}(t)$.  
We also find that the mean event duration does not depend simply on 
the barrier-top frequency (curvature), but is sensitive to details
of the potential.
All of the analytic results are confirmed by transition-path-sampling
simulations, implemented in a novel way.
\end{abstract}

\maketitle

\section{INTRODUCTION}
Stochastic descriptions of dynamics have been invoked to describe protein folding
~\cite{Naganathan-Munoz_Barrier_Heights_Protein_Folding-JACS-2005}, 
protein dynamics
~\cite{Schulten_Isralewitz-Steered_Molecular_Dynamics-COSB-2001}, 
chemical isomerization
~\cite{Zuckerman-Butane-JCP-2002}, 
and chemical reactions
~\cite{Smelyanskiy_Dykman-Nucleation_Rate_Periodically_Driven_System-JCP-1999}
among many other examples too numerous to list here.
A Brownian particle moving in a one-dimensional bistable potential typically provides a model 
for chemical and biological reaction systems
~\cite{Zwanzig-Nonequilibrium_Statistical_Mechanics-E1st-2001,
Berg-Random_Walks_in_Biology-E1st-1993,
Zuckerman-Important_Sampling-JCP-1999}.

\begin{figure}[H]
\begin{center}
\includegraphics{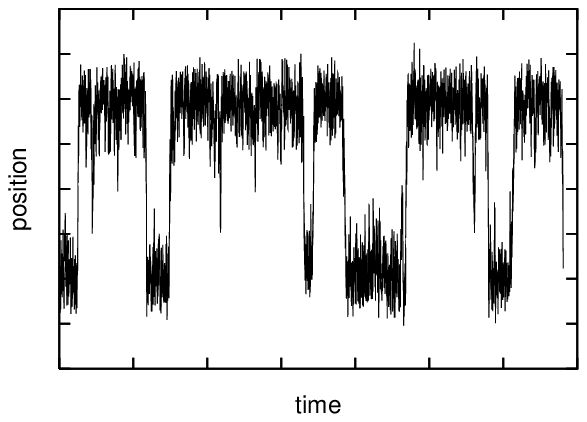}\includegraphics{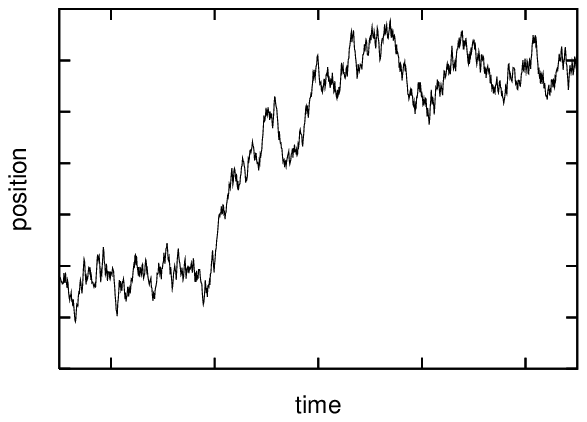}

\caption{\label{figINT1}Trajectory for a Brownian particle moving in a double 
well potential. The left graph is a long trajectory with several transition-events.
The right graph is the detail of a single transition-event cut from the same long trajectory.}
\end{center}
\end{figure}

\begin{figure}[H]
\begin{center}
\includegraphics{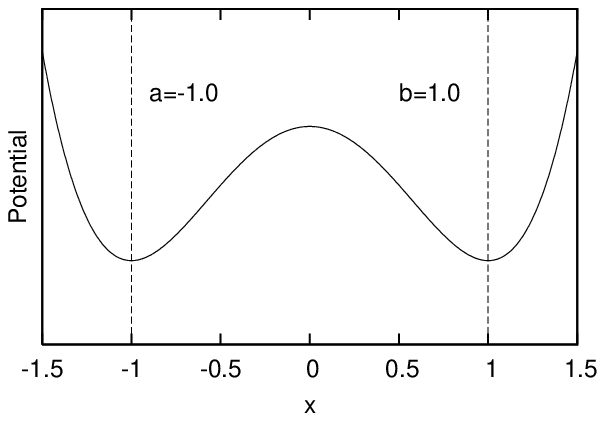}\includegraphics{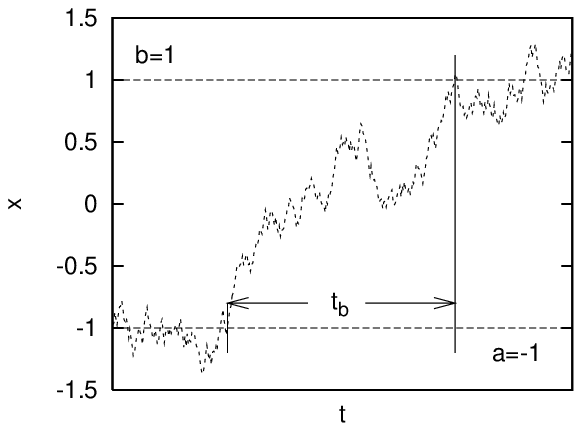}

\caption{\label{figINT2}The definition of transition-event duration. In the left graph, two
positions are defined as the start point $a$ and end point $b$ of the transition.
For a transition-event in the right graph, transition-event duration is the duration between last 
time the particle passes $a$  and the first time it passes $b$.}
\end{center}
\end{figure}

Basic aspects of the problem we address can be understood by examining Figs.~\ref{figINT1} 
and ~\ref{figINT2}, where we show a trajectory for a Brownian
particle moving in a one-dimensional double-well potential. There are two timescales of interest,
see
~\cite{Borkovec_Hanggi-Kramers_Law-RMP-1990}. 
One is the waiting time, or first passage time (FPT), which is the time the particle stays in one 
potential minimum before it
goes to the other minimum. The other timescale is that for just climbing over
the barrier separating the  two minima, excluding the waiting time.
We refer to this latter time as the ``transition-event duration''; it has also been termed the
``translocation time'' in the context of membrane and pore traversal
~\cite{Berezhkovskii_Hummer_Bezrukov-Equivalence_Trans_Paths_Ion_Channels-PRL-2006}.
In the right hand graph of Fig.~\ref{figINT1}, one transition-event extracted from the full time 
series on the left is shown at higher temporal resolution. See also Fig.~\ref{figINT2}.

Theoretical analysis of the first passage time is largely a textbook subject now
~\cite{Gardiner-Stochastic_Methods_Handbook-E2nd-1994, 
Risken-Fokker_Planck_Equation-E2nd-1989}.
In the simplest description, activated dynamics are modeled as Poisson processes, and the 
first passage time hence
follow an exponential distribution. The mean FPT is given by Kramers' theory
~\cite{Zwanzig-Nonequilibrium_Statistical_Mechanics-E1st-2001,
Borkovec_Hanggi-Kramers_Law-RMP-1990},
and depends on the height of the barrier as well as the curvatures at the 
minimum and at the barrier top. More complex models of the FPT distribution have also been 
studied
~\cite{Fisher_Kolomeisky-molecular_motor_force-PNAS-1999,
Margolin_Barkai-Reexamination_Kramers_approach-PRE-2005}.

Transition event durations have received focused attention only more recently
~\cite{Nelson_Lubensky-Polymer_Translocation_Through_Pore-BJ-1999,
Zuckerman-Butane-JCP-2002,
Berezhkovskii_Hummer_Bezrukov-Equivalence_Trans_Paths_Ion_Channels-PRL-2006,
Hajek_Alvarez-Trans_Paths_Ion_Channels-PRE-2006}.
The event duration is 
important because it reflects the detailed dynamics of an activated transition.
The statistics of event durations are of biological interest in many situations, such as 
transport in ion channels
~\cite{
Hummer-Nucleic_acid_transport-PNAS-2004,
Hajek_Alvarez-Trans_Paths_Ion_Channels-PRE-2006,
Berezhkovskii_Hummer_Bezrukov-Equivalence_Trans_Paths_Ion_Channels-PRL-2006}, 
polymer translocation through a pore
~\cite{Sung_Park-Polymer_Through_Hole-PRL-1996,
Nelson_Lubensky-Polymer_Translocation_Through_Pore-BJ-1999,
Muthukumar-Polymer_Through_Hole-JCP-1999,
Muthukumar-Polymer_Through_Hole-PRL-2001}.
Recently developed nanosecond and femtosecond-scale experimental techniques 
~\cite{Diller_Herbst-Femtosecond_Resolution-Science-2002,
Schultz_Samoylova-Femtosecond_Resolution-JACS-2005}
have the potential to probe, directly, transition events.
This short timescale is also fundamental to simulation techniques, such as path sampling
~\cite{Pratt-Path_Probability-JCP-1986,
Olender_Elber-Classical_Trajectories_With_Large_Time_Step-JCP-1996,
Chandler-Path_Sampling_02-JCP-1998,
Woolf-Path_Corrected_Functionals_Of_Stochastic_Trajectories-CPL-1998,
Elber_Olender-Stochastic_Path_Approach_for_Trajectories-JPCB-1999,
Zuckerman-Important_Sampling-PRE-2000,
Eastman_Doniach-Reaction_Path_Annealing-JCP-2001,
Bolhuis_Chandler-Transition_Path_Sampling-ARPC-2002};
see also 
~\cite{Berkowitz_McCammon_Northrup-Optimum_Reaction_Coordinate-JCP-1983,
Czerminski_Elber-Reaction-Path_Study_of_Conformational_Transitions-JCP-1990,
Choi_Elber-Reaction_Path_Study_Effect_of_Side_Chains-JCP-1991,
Fischer_Karplus-Conjugate_Peak_Refinement_for_Reaction_Paths-CPL-1992,
Huo_Straub-MaxFlux_Algorithm_for_Optimized_Reaction_Paths-JCP-1997,
Crehuet_Field-Temperature_Dependent_Nudged_Elastic_Band_Algorithm-JCP-2003,
Trygubenko_Wales-Doubly_Nudged_Elastic_Band_Method_JCP-2004}.
We further note that many previous theoretical treatments based on optimization of
the Onsager-Machlup action
~\cite{Onsager_Machlup-Fluctuations_and_Irreversible_Processes-PR-1953}
discuss aspects of the transition event duration
~\cite{Dykman_Krivoglaz-SP_JETP-1979,
Bray_McKane-Escape_Rate_Colored_Noise-PRL-1989,
Dykman-Fluctuational_Driven_by_Colored_Gaussian_Noise-PRA-1990,
McKane_Bray-Path_Integrals_and_Non_Markov_Processes_1-PRA-1990,
McKane_Bray-Path_Integrals_and_Non_Markov_Processes_2-PRA-1990,
McKane_Luckock-Path_Integrals_and_Non_Markov_Processes_3-PRA-1990,
McKane-Path_Integrals_and_Stationary_Probability_Distributions-JPA-1993,
Olender_Elber-Steepest_Descent_Path-JMST-1997,
Bier_Dean-Intrawell_Relaxation_of_Overdamped_Brownian_Particles-PRE-1999,
Elber_Shalloway-Temperature_Dependent_Reaction_Coordinates-JCP-2000}.
Also, the distribution of transition paths is directly described in the work of
Dykman and colleagues
~\cite{Dykman_McClintock-Optimal_Paths_and_Prehistory_Problem_for_Large_Fluctuations-PRL-1992,
Dykman_Luchinsky-Corrals_and_Critical_Behavior_of_Distribution_of_Paths-PRL-1996,
Dykman_Smelyanskiy-Distribution_of_Fluctuational_Paths-SM-1998}.

The transition-event duration may be defined in a simple way.
In particular, Fig.\ref{figINT2} shows that two
positions are defined as the start and end points of the transition; usually they are on the
different sides of a barrier. When a transition occurs, the event duration is the time 
interval between the last 
time the particle passes the start point and the first time it reaches the end point.
Although arbitrary choices are inherently necessary for the start and end points, 
these do not appear to affect the basic physics.

The probability distribution of transition-event durations was previously studied in a 
phenomenological way by Zuckerman and Woolf
~\cite{Zuckerman-Butane-JCP-2002}.
Hummer 
~\cite{Hummer-Transition_Path-JCP-2004}
gives an analytic formula for the mean transition-event duration for an arbitrary one-dimensional 
potential. Indirectly, Redner's study of the first passage time in an interval supplies important 
precedents for our work
~\cite{Redner-1st_Passage-2001}, 
as does Gardiner's book
~\cite{Gardiner-Stochastic_Methods_Handbook-E2nd-1994}.
Two groups have recently discussed the time-reversal symmetry of the $\rho_b$ distribution, 
albeit without attempting the detailed probe of the distribution itself
~\cite{Hajek_Alvarez-Trans_Paths_Ion_Channels-PRE-2006,
Berezhkovskii_Hummer_Bezrukov-Equivalence_Trans_Paths_Ion_Channels-PRL-2006},
which we pursue here. Other efforts directed at polymer translocation
~\cite{Sung_Park-Polymer_Through_Hole-PRL-1996,
Muthukumar-Polymer_Through_Hole-JCP-1999,
Muthukumar-Polymer_Through_Hole-PRL-2001}
investigated a related but distinct problem, critically differing in boundary conditions;
see Sec.~\ref{sec:transition_event}.

In this work, we first review the derivation of the probability distribution of the transition-event 
durations, $\rho_b(t)$, from the Fokker-Planck Equation (FPE) with particular boundary conditions. 
We then obtain novel results.
A recursive formula for all the moments of $\rho_b(t)$ is found, which permits accurate
numerical approximations of $\rho_b$ for an arbitrary potential. The short-time behavior of $\rho_b$ 
is studied by path integral techniques, yielding universal behavior along with a potential-dependent
correction. For a bistable potential with a high barrier (i.e. a ``double-well''), 
the long time behavior of $\rho_b(t\to \infty)$ will be discussed.

\section{Brownian Motion}

Our description of transition-events
will be based on the traditional approach of a one dimensional 
``reaction coordinate'' $x$
coupled to a thermal bath
~\cite{Zuckerman-Important_Sampling-PRE-2000}. 
The analysis assumes over-damped Brownian dynamics, which we will variously address
via a Langevin description, the associated Fokker-Planck equation
and related path integral methods. In the following three subsections
we introduce our notation, terminology, and the basic model.

The direct object of our study, the distribution of transition-event 
durations, will be fully introduced in Section~\ref{sec:transition_event}.

\subsection{Over-Damped Langevin Dynamics}

We consider Brownian, stochastic dynamics as governed by the over-damped
Langevin equation for a generalized coordinate $x$ in the presence of 
Gaussian white noise
~\cite{Zuckerman-Important_Sampling-PRE-2000},
\begin{equation}
\frac{dx}{dt}=\frac{F(x)}{\gamma}+R(t)\,,
\label{eq:OD1}
\end{equation}
where 
\begin{equation}
F(x)=-\frac{\partial U(x)}{\partial x}\,.
\label{eq:OD1.1}
\end{equation}
is the physical, systematic, conservative force acting on the particle, based
on the potential energy $U$, and $\gamma$ is the friction constant. In this
work, the noise, $R(t)$, is taken to be Gaussian and white with zero mean and
correlation 
\begin{equation}
\langle R(t)R(t')\rangle=\left(\frac{2k_{B}T}{\gamma}\right)\delta(t-t')\,.
\label{eq:OD2}
\end{equation}
Here $k_{B}$ is Boltzmann's constant, and $T$ is the temperature.

Individual realizations of the noise in Eq.(\ref{eq:OD1}) generate
stochastic trajectories $x(t)$, which are routinely simulated numerically
as described in Section~\ref{sec:SIMULATION}.

\subsection{Fokker-Planck Equation}

To extract statistical information on trajectories one generally turns
from the Langevin equation to the associated Fokker-Planck equation (FPE).
The relation is discussed in standard monographs; see, for example
~\cite{Zwanzig-Nonequilibrium_Statistical_Mechanics-E1st-2001}.
The FPE describes the average behavior of a statistical ensemble of 
trajectories $x(t)$.

In one dimension the time evolution of the probability density function
$P(x,t)$, for the coordinate $x$ is assumed to be described by the
Fokker-Planck equation
~\cite{Gardiner-Stochastic_Methods_Handbook-E2nd-1994}
\begin{equation}
\frac{\partial P(x,t)}{\partial t}=\left[-\frac{\partial}{\partial x}D^{(1)}(x)+\frac{\partial^{2}}{\partial x^{2}}D^{(2)}(x)\right]P(x,t)\,.
\label{eq:FPE1}
\end{equation}
In Eq.(\ref{eq:FPE1}), $D^{(2)}(x)>0$ is the diffusion coefficient
and $D^{(1)}(x)$, is the drift coefficient. When $D^{(2)}={k_{B}T}/{\gamma}\equiv D$,
with $D$ the diffusion constant, and $D^{(1)}(x)=F(x)/\gamma$,
this Fokker-Planck equation embodies the over-damped Langevin dynamics 
of Eq.(\ref{eq:OD1}) with noise correlation satisfying Eq.(\ref{eq:OD2})
~\cite{Zwanzig-Nonequilibrium_Statistical_Mechanics-E1st-2001}.
In this simplified case we obtain 
\begin{equation}
\frac{\partial P(x,t)}{\partial t}=-\frac{\partial}{\partial x}\left\{ -D\left[\frac{dU^{\star}(x)}{dx}+\frac{\partial}{\partial x}\right]P(x,t)\right\} \,,
\label{eq:FPE2}
\end{equation}
where $U^{\star}(x)={U(x)}/{k_{B}T}$ is the dimensionless physical
potential. The solution of the Fokker-Planck equation under suitable
initial and boundary conditions will allow statistical information
to be extracted.

Eq.(\ref{eq:FPE2}) clearly expresses conservation of probability.
The total current associated with the stochastic variable $x$ is
given by
~\cite{Gardiner-Stochastic_Methods_Handbook-E2nd-1994}
\begin{equation}
J(x,t)=-D\left[\frac{dU^{\star}(x)}{dx}+\frac{\partial}{\partial x}\right]P(x,t)\,,
\label{eq:FPE3}
\end{equation}
where the first term represents the drift (systematic) contribution,
while the second term is the diffusion contribution.

\subsection{Path Integral Approach\label{sec:Path_Integral}}

Path integral methods provide a useful tool
and a different perspective for the study of Brownian motion
~\cite{Wiegel-Path_Integral-1986,
Kleinert-Path_Integral-2004,
Schulman-Path_Integral-1st-1981}. 
Trajectories $x(t)$ are directly considered, rather than their average as in the FPE.
The relative probability of a trajectory, $W[x(t)]$, connects the path integral back to the
original Langevin equation (\ref{eq:OD1}).
It can be shown that the relative probability for a Brownian particle
described by the Langevin equation (\ref{eq:OD1}) to follow
a specific path $x(\tau)$, with $\tau$ the time, is given by
~\cite{Wiegel-Path_Integral-1986}
\begin{equation}
W[x(\cdot)]=\exp\left\{ -\frac{1}{4D}\int_{t_{0}}^{t}L[x(\tau)]d\tau\right\} \,,
\label{eq:PI1}
\end{equation}
where the effective Lagrangian can be expressed as 
\begin{equation}
L[x(\tau)]=\left(\frac{dx}{d\tau}-\frac{F}{\gamma}\right)^{2}+\frac{2D}{\gamma}\frac{dF}{dx}\,.
\label{eq:PI2}
\end{equation}
Then the optimal (classical) path $x_{c}(\tau)$ can be found in
the standard fashion from the stationarity of the exponent ``action'',
\begin{equation}
\delta\int_{t_{0}}^{t}L[x(\tau)]d\tau=0\,,
\label{eq:PI3}
\end{equation}
with $x(t_{0})=a$ and $x(t)=x$. Integrating over the probability densities
of individual paths, the propagator 
\begin{equation}
G(x,t|a,t_{0})=\int_{a,t_{0}}^{x,t}W[x(\tau)]d[x(\tau)]\,,
\label{eq:PI4}
\end{equation}
determines the probability of the particle arriving at position $x=x(t)$
at time $t$ given that it started at $a=x(t_{0})$. With a suitable definition of
the measure in the integration in (\ref{eq:PI4}) the propagator
is equivalent to the {} ``principal solution'' or Green function
solution of the associated Fokker-Planck equation
~\cite{Wiegel-Path_Integral-1986,
Schulman-Path_Integral-1st-1981}.


When the diffusion coefficient $D$ is small, the major contribution
to the propagator will come from the paths very close to the optimal
path $x_c(\tau)$. So, for $D\rightarrow0$, the simplest approximation retains
only the contribution from the optimal path
~\cite{Wiegel-Path_Integral-1986}, 
and is of the form 
\begin{equation}
G(x,t|a,t_{0})\cong K(t)\exp\left\{ -\frac{1}{4D}\int_{t_{0}}^{t}L[x_{c}(\tau)]d\tau\right\} \,,
\label{eq:PI5}
\end{equation}
where $K(t)$ is the normalizing factor to ensure the propagator
satisfies 
\begin{equation}
\int_{-\infty}^{\infty}G(x,t|a,t_{0})dx=1\,.
\label{eq:PI6}
\end{equation}

To include small fluctuations around the classical path, one typically
invokes a quadratic approximation, in which deviations
to second order are retained in the effective Lagrangian. Writing
\begin{equation}
x(\tau)=x_{c}(\tau)+\delta(\tau)\,,
\label{eq:PI7}
\end{equation}
with $\delta(t_0)=\delta(t)=0$, and neglecting terms in the Lagrangian
higher than the second order in $\delta$, one finds
~\cite{Wiegel-Path_Integral-1986}
\begin{eqnarray}
G(x,t|a,t_{0}) & = & \exp\left(\frac{1}{2k_{B}T}\int_{a}^{x}Fdx'\right)\left\{ 4\pi D[2E+4DV(a)]^{\frac{1}{2}}[2E+4DV(x)]^{\frac{1}{2}}\int_{a}^{x}(2E+4DV(x'))^{-\frac{3}{2}}dx'\right\} ^{-\frac{1}{2}}\nonumber \\
 & \times & \exp\left[-\frac{1}{4D}\int_{t_{0}}^{t}\left(\frac{dx_{c}}{d\tau}\right)^{2}d\tau-\int_{t_{0}}^{t}V(x_{c})d\tau\right]\,.
\label{eq:PI8}
\end{eqnarray}
Here the effective potential, $V(x)$, (distinct from the physical
potential $U$) is given by 
\begin{equation}
V(x)=\frac{F(x)^{2}}{4k_{B}T\gamma}+\frac{1}{2\gamma}\frac{dF}{dx}\,.
\label{eq:PI9}
\end{equation}
The constant $E$ in Eq.(\ref{eq:PI8}) is the ``energy'' of the particle, appearing
as an integration constant in the first integral of the effective
equation of motion following from the Lagrangian in Eq.(\ref{eq:PI2})
and the extremization in Eq.(\ref{eq:PI3}). The first integral yields
the equation of motion, 
\begin{equation}
\frac{dx_{c}}{d\tau}=\{2E+4DV[x_{c}(\tau)]\}^{\frac{1}{2}}\,,
\label{eq:PI10}
\end{equation}
which is the equation for a classical particle of mass $\frac{1}{2}$
in the presence of a conservative potential $-2DV(x)$
~\cite{Wiegel-Path_Integral-1986}.
Larger values of $E$ give optimal solutions for increasingly more 
rapid events.

The quadratic approximation is justified if the path probability decreases sufficiently rapidly
with increasing variation of the path from the optimal one
~\cite{Wiegel-Path_Integral-1986}.
This is expected in the limit of weak diffusion $(D \to 0)$ in analogy with the semi-classical
approximation in quantum mechanics with $\hbar \to 0$; see,e.g.,
~\cite{Wiegel-Path_Integral-1986}.

The path integral approach provides some interesting insights for
the features of Brownian dynamics of concern in this paper. We will
return to this description below.

\section{SIMULATION\label{sec:SIMULATION}}
All of our key analytic results to be discussed below have been confirmed via numerical simulation.
Here our simulation approaches are briefly described.

\subsection{\label{sub:Simulation-by-Langevin}Brute Force Simulation}
Standard simulations of the over-damped Langevin Eq.(\ref{eq:OD1}) employ a simple first-order 
scheme with fixed time step $\Delta t$, such that
\begin{equation}
x_{j}=x(j\Delta t),\ \ j=0,1,2...
\end{equation}
and
~\cite{Zuckerman-Important_Sampling-JCP-1999}
\begin{equation}
x_{j+1}=x_{j}+\frac{F(x_{j})}{\gamma}\Delta t+\Delta x_{R}\,.
\label{eq:SLE1}
\end{equation}
Consistent with Eq.(\ref{eq:OD2}) the thermal fluctuation (noise increment) 
$\Delta x_{R}$ is chosen from a Gaussian 
distribution of zero mean and variance
\begin{equation}
\sigma^{2}=2\left(\frac{k_{B}T}{\gamma}\right)\Delta t=2D\Delta t\,.
\label{eq:SLE2}
\end{equation}

However, as is well-known, this direct approach proves inadequate to simulate rare events, even
in one dimension. A program running on a single CPU can
provide an ensemble of transition trajectories (with thousands of transition-events) only for low barrier
height. For high barriers, the waiting time between
successful events will become unacceptably long. Therefore, we employed a
path-sampling method for simulations, which we now describe.

\subsection{\label{sec:path-sampling}Path Sampling}

The problem with direct ``brute force'' simulations 
is that the waiting time between events grows 
exponentially with barrier height
~\cite{Risken-Fokker_Planck_Equation-E2nd-1989}.
Our interest here, moreover, is to obtain a statistically 
well-sampled ensemble of transitions. In practice, for any given model, we 
require thousands of events.

To generate a sufficient quantity of transition-events, we turn to a Monte Carlo path-sampling 
approach. The approach has its roots in Path Integral Monte Carlo for quantum systems
~\cite{Ceperley_Pollock-Path_Integral_MC_QM-PRB-1984,
Ceperley_Pollock-Path_Integral_MC_4He-PRL-1986}, 
but Pratt provided an important advance in recognizing the analogous application 
in classical and, particularly, chemical systems
~\cite{Pratt-Path_Probability-JCP-1986}.

Pratt's approach has recently been taken up with some vigor by Chandler and co-workers
~\cite{Chandler-Path_Sampling_01-JCP-1998,
Chandler-Path_Sampling_02-JCP-1998}.
Related work was presented by Zimmer and Paniconi
~\cite{Zimmer-MONTE_CARLO_SPACE_TIME_CONFIGURATIONS-PRL-1995,
Paniconi_Zimmer-Large_Fluctuations-PER-1999}.
An independent path-sampling approach was developed by Zuckerman and Woolf
~\cite{Woolf-Path_Corrected_Functionals_Of_Stochastic_Trajectories-CPL-1998,
Zuckerman-Important_Sampling-JCP-1999}, 
building on work by Ottinger
~\cite{Ottinger-Variance_Reduced_BD-MM-1994}.

The basic idea of path-sampling is simple: focus computer time on the rare 
transition events of 
interest, Fig.~\ref{figINT1}(right), rather than on the waiting time between events, which can be
longer by many orders of magnitude, Fig.~\ref{figINT1}(left). In a statistical mechanics 
context, the probability of 
a path can be computed. Hence ``trial'' paths can be included by means of re-weighting
~\cite{Zuckerman-Important_Sampling-JCP-1999}
or by a Metropolis criterion
~\cite{Chandler-Path_Sampling_02-JCP-1998}.

Here we primarily follow Pratt's approach to path sampling, which is based on two facts:
(i) Path (i.e., trajectory) probabilities are readily computed for stochastic processes,
so that trajectories may be viewed as $N\times d$ dimensional equilibrium ``objects'', when there
are $N$ time steps and $d$ spatial dimensions.
(ii) Wherever equilibrium probabilities can be computed for all such ``objects'' in a space, 
Metropolis sampling can be performed.

As in any Metropolis simulation, we require that detailed balance is satisfied. That
is, for arbitrary paths $i$ and $j$ with equilibrium probabilities $P$ and overall
transition rates $\Gamma$, we require 
\begin{equation}
P_{path}(i)\Gamma(i \to j)=P_{path}(j)\Gamma(j \to i)\,.
\label{eq:SPS.0.1}
\end{equation}
The rate $\Gamma$ is decomposed into the usual product of the 
generating ($gen$) and acceptance ($acc$) components
\cite{Frenkel_Smit-Molecular_Simulation-E2nd-2001,
Chandler-Path_Sampling_02-JCP-1998}, 
which are proportional to the conditional probability for generating and
accepting the trial path $j$, starting from $i$. Then trial moves should be accepted 
with probability $\min[1,R]$, where
\begin{equation}
R=\frac{acc(i \to j)}{acc(j \to i)}=\frac{P_{path}(j) \times gen(j \to i)}{P_{path}(i) \times gen(i \to j)}\,.
\label{eq:SPS.0.2}
\end{equation}

All paths in our ensemble will have the same total number of steps $N$, so that
the probability of two paths can be compared via Eq.(\ref{eq:SPS.0.2}). We will typically choose 
$N$ to be much bigger than $\langle t \rangle_{b}/\Delta t$, with $\langle t \rangle_{b}$ the
mean time for transition events, so that the ``full shape'' of the distribution is sampled.
Thus, transition events will typically constitute only part of $N$-step trajectories in our
sample of paths. On the other hand, intentionally selecting $N$ smaller than 
$\langle t \rangle_{b}/\Delta t$ when necessary allows us to focus on the short-time behavior of $\rho_b$.

To proceed, we must establish the equilibrium and generating probabilities in Eq.(\ref{eq:SPS.0.2}).
The ``equilibrium'' probability $P_{path}$ of the $N$-step path from $a$ to $x_{N}$ is the product of the 
equilibrium probability for the initial point and all subsequent  
single-step transition probabilities 
consistent with Eqs.(\ref{eq:SLE1}) and (\ref{eq:SLE2}). We further restrict our ensemble 
to ``successful'' paths containing transition events by formally introducing 
a projection operator $\theta$. 
Thus we have
\begin{equation}
P_{path}(\{a,x_1,...,x_N\})\propto \exp[-U^{\star}(a)]\times[\Pi_{i=0}^{N-1}p(x_{i},x_{i+1};U^{\star})]\times\theta(\{x_i\})\,.
\label{eq:SPS.1}
\end{equation}
The single-step transition probability corresponding to Eqs.(\ref{eq:SLE1}) and (\ref{eq:SLE2}) 
is a Gaussian density, namely
\begin{equation}
p(x_{i},x_{i+1};U^{\star})=\frac{1}{\sqrt{2\pi}\sigma}\exp\left\{ -\frac{\left[x_{i+1}-x_{i}-\frac{1}{2}\left(\frac{dU^{\star}}{dx_i}\right)(2D\Delta t)\right]^{2}}{2\sigma^{2}}\right\}\,.
\label{eq:SPS.2}
\end{equation}
where $\frac{dU^{\star}}{dx_i}\equiv\left.\frac{dU^{\star}(x)}{dx}\right|_{x_i}$.
If the particle returns to the left boundary $a$ before arriving at 
the right boundary $b$, $\theta=0$; otherwise $\theta=1$.

We employ a novel path generating procedure designed to focus simulation effort on the
distribution $\rho_{b}(t)$ of interest and, as necessary, on the rarest trial events.
Our path generation strategy is closely related to a non-Metropolis 
re-weighting procedure previously considered by Zuckerman 
and Woolf 
~\cite{Zuckerman-Important_Sampling-JCP-1999}.
Specifically, a trial path is built up ``from scratch'', but based 
on the average behavior of the previous path.
From the previous path, which starts from $a$ and arrives at $b$
after $M_{old}$ ($M_{old}<N$) steps without being absorbed at $a$, we can calculate 
the average velocity over total time $M_{old}(2D\Delta t)$ as
\begin{equation}
\bar{v}(M_{old})=\frac{b-a}{M_{old}(2D\Delta t)}\,.
\label{eq:SPS.3}
\end{equation}
This will be the ``target speed'' of the new trial path. This is extremely useful
when studying the fastest events, whose transition-event durations are much shorter than
$\langle t \rangle_{b}$.

To generate a new path, we linearly bias the particle from $a$ to $b$ using
\begin{equation}
x_{j+1}=x_j+(\bar{v}(M_{old}))(2D\Delta t)+\Delta x_{R}\,,
\label{eq:SPS.4}
\end{equation}
where $\Delta x_{R}$ has been defined in Section~\ref{sub:Simulation-by-Langevin} following 
Eq(\ref{eq:SLE1}). Eq.(\ref{eq:SPS.4}) may be compared to the unbiased form (\ref{eq:SLE1}). 
The linear bias in (\ref{eq:SPS.4}) is motivated by the quasi-ballistic quality of the fastest 
transition-events deriving from Eq.(\ref{eq:PI10}) in the limit $E \gg DV_{max}$, where $V_{max}$ 
is the maximum of $V(x)$ defined by Eq.(\ref{eq:PI9}).
Thus, on the new path, the particle moves with a constant drift (bias) velocity, as if the
force were constant, and ordinary noise.
Note that for the new trajectory generated by Eq.(\ref{eq:SPS.4}), the new value, $M_{new}$, can be 
larger or smaller than $M_{old}$. Once it arrives at $b$, we remove the bias and allow the particle to 
move for the remainder of $N$ steps as governed by unbiased Brownian motion Eq.(\ref{eq:SLE1}). As
noted above, all the paths must contain the same number of steps for probabilities to be
compared in our Metropolis procedure using Eq.(\ref{eq:SPS.0.2})

The generating probability ($gen$) for our procedure is the conditional probability with which we
choose the new path, given the old one (with its average speed), namely,
\begin{equation}
gen(old\rightarrow new)=\prod_{i=0}^{M_{new}-1}\bar{g}(x_{i},x_{i+1};\bar{v}(M_{old}))\prod_{j=M_{new}}^{N-1}p(x_{j},x_{j+1};U^\star)\,,
\label{eq:SPS.5}
\end{equation}
where 
\begin{equation}
\bar{g}(x_{i},x_{i+1},\bar{v}(M_{old}))=\frac{1}{\sqrt{2\pi}\sigma}\exp\left\{-\frac{[x_{i+1}-x_{i}-(\bar{v}(M_{old}))(2D\Delta t)]^{2}}{2\sigma^{2}}\right\}\,.
\label{eq:SPS.6}
\end{equation}
This generating method is tailored to the potentials and boundary conditions
we study in this paper, so that the Brownian particle will not be trapped in any 
position between the two absorbing walls. 

By substituting Eqs.(\ref{eq:SPS.1}),(\ref{eq:SPS.2}),(\ref{eq:SPS.5}) and (\ref{eq:SPS.6})
into Eq.(\ref{eq:SPS.0.2}) we arrive at the acceptance criterion for our
generating procedure, namely 
\begin{equation}
R=\frac{\prod_{i=0}^{M_{new}-1}p(x_{i},x_{i+1},U^{\star})\theta_{new}\prod_{i=0}^{M_{old}-1}\bar{g}(y_{i},y_{i+1},\bar{v}(M_{new}))}{\prod_{i=0}^{M_{new}-1}\bar{g}(x_{i},x_{i+1},\bar{v}(M_{old}))\theta_{old}\prod_{i=0}^{M_{old}-1}p(y_{i},y_{i+1},U^{\star})}\,,
\label{eq:A.3.4}
\end{equation}
given an old transition path $(a,y_{1},y_{2}......y_{M_{old}},y_{M_{old}+1}.....y_{N})$
and a trial transition path $(a,x_{1},x_{2}......x_{M_{new}},x_{M_{new}+1}.....x_{N})$.

To ensure the correct behavior  of our procedure and code,
path-sampling results were carefully checked against direct simulation, using Eq.(\ref{eq:SLE1}), 
in a number of cases. In the following sections of the paper, all the simulation results employ
the path-sampling method just described.

We also checked that our path-sampling simulations greatly exceeded the correlation time 
resulting from our use of the ``old'' average velocity in Eq.(\ref{eq:SPS.3}). This resulted 
in negligible statistical uncertainty, as can be gauged from the smoothness of the data
in all path-sampling figures.

\section{Transition-Events\label{sec:transition_event}}

\subsection{Distribution of Transition-Event Duration\label{sec:Trans_Event}}

The distribution of transition-event durations, $\rho_b(t)$, for a Brownian particle
confined to one dimension can be found by solving the 
Fokker-Planck equation using suitable boundary conditions, as we now describe
~\cite{Nelson_Lubensky-Polymer_Translocation_Through_Pore-BJ-1999}.

During the entire transition process the particle must move between
$a$ and $b$, which means that only trajectories that stay completely
within the interval are considered, i.e. $a \le x(t) \le b$ during 
the entire event. To eliminate the extraneous trajectories,
absorbing walls must be put at the start and end points,
~\cite{Nelson_Lubensky-Polymer_Translocation_Through_Pore-BJ-1999}, 
so that
\begin{eqnarray}
P(a,t) & = & 0\nonumber \\
P(b,t) & = & 0\,.
\label{eq:DTE1}
\end{eqnarray}
As recently stressed by Berezhkovskii 
\cite{Berezhkovskii_Hummer_Bezrukov-Equivalence_Trans_Paths_Ion_Channels-PRL-2006},
et al., the dual absorbing boundary conditions distinguish
the event duration as a ``conditional first passage time'', rather than the usual unconditional
time associated with the Kramers' problem
~\cite{Borkovec_Hanggi-Kramers_Law-RMP-1990,
Berezhkovskii_Hummer_Bezrukov-Equivalence_Trans_Paths_Ion_Channels-PRL-2006}.
This contrasts with several previous studies of polymer translocation
~\cite{Berezhkovskii_Hummer_Bezrukov-Equivalence_Trans_Paths_Ion_Channels-PRL-2006,
Muthukumar-Polymer_Through_Hole-JCP-1999,
Muthukumar-Polymer_Through_Hole-PRL-2001}.

One releases particles very close to the left absorbing wall at
$t=0$, so that the initial condition is 
\begin{equation}
P(x,0)=\delta[x-(a+\epsilon)]\,,
\label{eq:DTE3}
\end{equation}
with $\epsilon\rightarrow0+$. Then the current at the right absorbing
wall will determine the distribution, $\rho_{b}$, of durations according to
\begin{equation}
\rho_{b}(t)\,\,\propto\lim_{\epsilon\rightarrow0+}J(b,t)\,,
\label{eq:DTE4}
\end{equation}
with the currents given in Eq.(\ref{eq:FPE3}).
Following Gardiner's work
~\cite{Gardiner-Stochastic_Methods_Handbook-E2nd-1994}, 
let $\pi_{b}(a+\epsilon|t)$ equal the probability that
a particle, released at $a+\epsilon$, is absorbed at the
right absorbing wall during $0<\tau<t$. It is easy to see that
\begin{equation}
\pi_{b}(a+\epsilon|t)=\int_{0}^{t}J(b,\tau)d\tau\,.
\label{eq:DTE5}
\end{equation}
If we define 
\begin{equation}
\Pi_{b}(a+\epsilon)\equiv\pi_{b}(a+\epsilon|\infty)=\int_{0}^{\infty}J(b,\tau)d\tau\,,
\label{eq:DTE6}
\end{equation}
this ``splitting probability'' can be used to normalize $\rho_{b}(t)$
in Eq.(\ref{eq:DTE4}) according to
\begin{equation}
\rho_{b}(t)=\lim_{\epsilon\rightarrow0+}\frac{J(b,t)}{\int_{0}^{\infty}J(b,\tau)d\tau}=\lim_{\epsilon\rightarrow0+}\frac{J(b,t)}{\Pi_{b}(a+\epsilon)}\,.
\label{eq:DTE7}
\end{equation}
We note that the splitting probabilities are time-independent and follow directly from
the potential $U^\star$ according to
~\cite{Gardiner-Stochastic_Methods_Handbook-E2nd-1994, Hummer-Transition_Path-JCP-2004}:
\begin{eqnarray}
\Pi_{a}(x) & = & \frac{\int_{x}^{b}\exp[U^{\star}(x')]dx'}{\int_{a}^{b}\exp[U^{\star}(x')]dx'}\nonumber \\
\Pi_{b}(x) & = & \frac{\int_{a}^{x}\exp[U^{\star}(x')]dx'}{\int_{a}^{b}\exp[U^{\star}(x')]dx'}=1-\Pi_{a}(x)\,.
\label{eq:DTE8}
\end{eqnarray}

Hence, to find the distribution of the transition-event durations, $\rho_{b}(t)$,
one must solve the Fokker-Planck equation (\ref{eq:FPE2}) with
the initial condition (\ref{eq:DTE3}) and absorbing boundary conditions
(\ref{eq:DTE1}). The current, $J(b,t)$,
can be found from Eq.(\ref{eq:FPE3}), which can then be combined with
the splitting probabilities to find the normalized distribution of transition-event
durations, $\rho_{b}(t)$.

\subsection{Examples: Free Diffusion and Linear Potential}

The solution of the Fokker-Planck equation can be formally expressed, in standard
fashion, in terms of the eigenvalues and eigenfunctions of a time independent 
equation~
\cite{Risken-Fokker_Planck_Equation-E2nd-1989}. 
The solution can be written in the form
\begin{equation}
P(x,t)=\sum_{n}A_{n}p_{n}(x)e^{-\lambda_{n}t}\,,
\label{eq:EM1}
\end{equation}
where the eigenvalues $\{\lambda_{n}\}$ are non-negative and, based on Eq.(\ref{eq:FPE2}), 
the eigenfunctions satisfy
\begin{equation}
D\left\{ \frac{d}{dx}\left[\frac{dU^{\star}(x)}{dx}\right]+\frac{d^{2}}{dx^{2}}\right\} p_{n}(x)=-\lambda_{n}p_{n}(x)\,.
\label{eq:EM2}
\end{equation}
Eq.(\ref{eq:EM2}) with boundary conditions (\ref{eq:DTE1}) determine the functions $\{ p_{n}(x)\}$, 
while the constants
$\{ A_{n}\}$ are found from the initial condition (\ref{eq:DTE3}). The distribution $\rho_{b}(t)$ follows 
from Eqs.(\ref{eq:DTE7}) and (\ref{eq:DTE8}). As examples, we determine $\rho_{b}(t)$ for a few special
potentials $U^{\star}(x)$. This will reveal some interesting features. We note that the linear potential,
of which free diffusion is a special case, previously was studied by Lubensky and Nelson
~\cite{Nelson_Lubensky-Polymer_Translocation_Through_Pore-BJ-1999},
although without numerical simulations.

\emph{Free Diffusion.} Even in the absence of a true barrier, the event duration is still
well defined by the formalism above, and this simple case acts as a useful
reference. We therefore first consider free diffusion, with $U^\star(x)=0$ and $a=0$, 
$b=L$.

The solution of Eq.(\ref{eq:EM2}) can easily be found, and the result can be formally expressed as
\begin{equation}
\rho_{b}^{(0)}(t)=2D\sum_{j=1}^{\infty}(-1)^{j+1}\left(\frac{j\pi}{L}\right)^{2}\exp\left(-\frac{j^{2}\pi^{2}}{L^{2}}Dt\right)\,.
\label{eq:FU1}
\end{equation}
Notice that at long times the decay is exponential and dominated by the lowest eigenvalue.

The right-hand side of Eq.(\ref{eq:FU1}) is well behaved for long time, but is not useful for 
$t \to 0$. We can re-cast the result in a format useful at short times by using the Poisson sum 
formula
~\cite{Poisson_Sum-MathWorld},
\begin{equation}
\sum_{n=-\infty}^{\infty}f(n)=\sum_{j=-\infty}^{\infty}\int_{-\infty}^{\infty}f(x)\exp(-2i \pi jx)dx\,,
\label{eq:FU1.1}
\end{equation}
for function $f$.
We then find an alternative representation
\begin{equation}
\rho_{b}^{(0)}(t)=\frac{2L}{\sqrt{\pi Dt^3}}\sum_{j=0}^{\infty}\left[\frac{(2j+1)^2L^2-2Dt}{4Dt}\right]\exp\left[-\frac{(2j+1)^2L^2}{4Dt}\right]\,,
\label{eq:FU1.2}
\end{equation}
which can be used to extract the behavior as $t \to 0$, namely $\rho_{b}^{(0)}(t) \sim t^{-5/2}\exp[-L^2/(4Dt)]$. 
We note that Eq.(\ref{eq:FU1.2}) can also be derived using
an image method, as described below in Section~\ref{sec:free_diff}.

\emph{Linear Potential.} The solution for the linear potential, $U^{\star}(x)=kx$, which corresponds to a constant
drift velocity, can also be formally written in terms of an eigenfunction expansion
~\cite{Nelson_Lubensky-Polymer_Translocation_Through_Pore-BJ-1999},
\begin{equation}
\rho_{b}(t)=2D\sum_{j=1}^{\infty}(-1)^{j+1}\left(\frac{j\pi}{L}\right)^{2}\left[\frac{\sinh(kL/2)}{kL/2}\right]\exp\left\{ -\frac{[(kL/2)^{2}+j^{2}\pi^{2}]}{L^{2}}Dt\right\}\,.
\label{eq:LU1}
\end{equation}
Comparing with Eq.(\ref{eq:FU1}), the result can be written
\begin{equation}
\rho_{b}(t)=\rho_{b}^{(0)}(t)\left[\frac{\sinh(kL/2)}{kL/2}\right]\exp\left[-\frac{(kL/2)^{2}}{L^{2}}Dt\right]\,.
\label{eq:LU2}
\end{equation}

\begin{figure}[H]
\begin{center}
\includegraphics{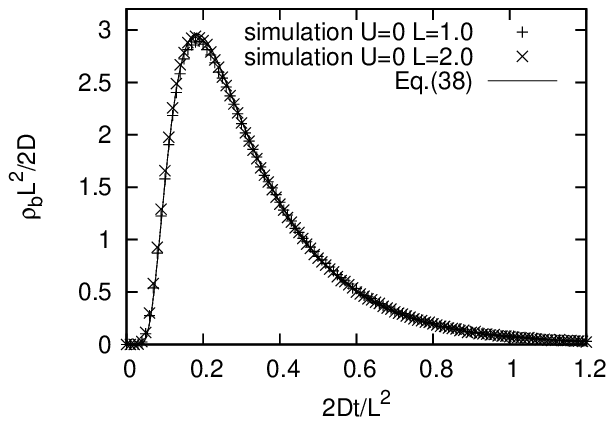}\includegraphics{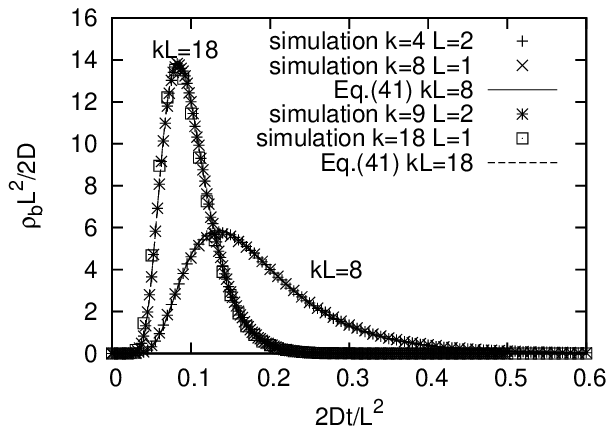}

\caption{\label{figSFLP}Scaled event duration distribution for free diffusion and the linear 
potential. The left graph is for free diffusion with different 
widths $L$, where data points are from path-sampling simulations and numerical evaluations 
of Eq.(\ref{eq:FU1}).
The right graph is for the potential $U^{\star}=kx$ with different widths $L$ and different 
slopes $k$; the points
are data from path-sampling simulations and the lines are numerical evaluations of Eq.(\ref{eq:LU1}). 
}

\end{center}
\end{figure}

In the left graph of Fig.~\ref{figSFLP} we show path-sampling simulation
results following Section~\ref{sec:path-sampling} 
for $\rho_{b}$ for free diffusion with $U^\star=0$, $L=1.0$ and $U^\star=0$, $L=2.0$.
They are compared with the numerical evaluations of  
Eq.(\ref{eq:FU1}). 
The path-sampling simulations and numerical results from 
the eigenfunction expansions match very well. 
We changed the units of the vertical and horizontal axes, so that all the
curves of $\rho_b^{(0)}$ will not depend on the width $L$, and the generic behavior is highlighted.

In the right graph of Fig.~\ref{figSFLP} we show path-sampling simulation
results for the a series of parameterizations of linear potential: 
$U^\star=4.0x$, $L=2.0$, $U^\star=8.0x$, $L=1.0$,
$U^\star=9.0x$, $L=2.0$, and $U^\star=18.0x$, $L=1.0$, which are compared
to numerical evaluation of Eq.(\ref{eq:LU1}). The simulation 
and numerical results again match very well. 
We again scaled the axes to emphasize that the shape of
$\rho_b$ only depends on the value of $kL$, which is essentially the potential energy
difference between the start and end points. These exercises add confidence to the 
path-sampling methods used here.

\subsection{Approximate Solution for Inverted Parabolic Potential }
\begin{figure}[H]
\begin{center}
\includegraphics{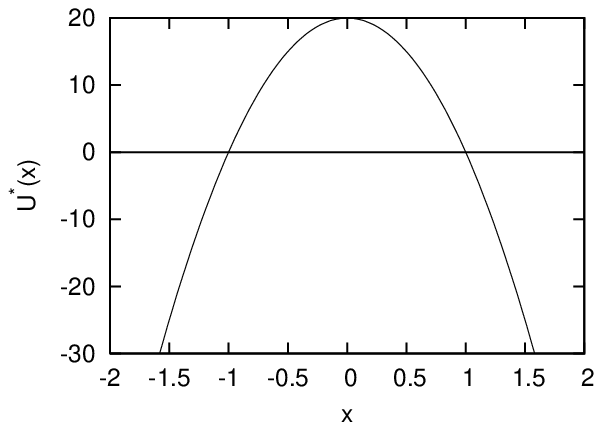}\includegraphics{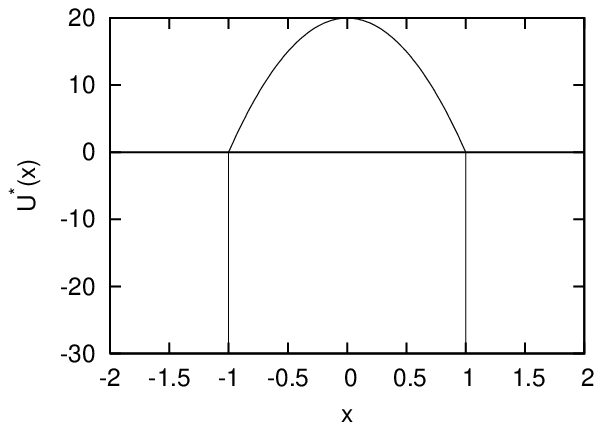}

\caption{\label{figIPP1}Inverted parabolic potential with different boundary conditions.
The left graph is the inverted parabolic potential $U^{\star}=-20(1-x^2)$,
with open boundary condition. The right graph is the same potential with two absorbing walls at
$x=-1$ and $x=1$.}
\end{center}
\end{figure}

As a first investigation of a more realistic potential, we employ a crude representation of 
absorbing boundary conditions. In Fig.~\ref{figIPP1}, inverted parabolic potentials are shown, one
with open boundary conditions and the other with two absorbing walls($U^\star\to-\infty$;
see, e.g.,
~\cite{Risken-Fokker_Planck_Equation-E2nd-1989}). 
When the ``barrier'' is high i.e., when $U^{\star}(0) \gg U^{\star}(a=-1) = U^{\star}(b=1)$, 
a particle exiting the region $a<x<b$, has a small likelihood of returning 
with open boundary conditions, because of the rapidly increasing ``downhill'' forces
external to the region. Thus, as long as there is a sufficiently high 
barrier, one might conclude the solution for open boundary conditions will be a good 
approximation for an inverted parabolic potential with two absorbing walls. We now investigate 
this approximation.

With open boundary conditions, the exact solution of the Fokker-Planck equation for an inverted
parabolic potential, $U^\star=-\frac{1}{2}\alpha x^2$, is well known
~\cite{Risken-Fokker_Planck_Equation-E2nd-1989}:
\begin{equation}
P(x,t)=\sqrt{\frac{\alpha}{2\pi[1-\exp(-2\alpha Dt)]}}\exp\left\{ -\frac{\alpha[x\exp(-\alpha Dt)-a]^{2}}{2[1-\exp(-2\alpha Dt)]}\right\} \exp(-\alpha Dt)\,,
\label{eq:IPP1}
\end{equation}
which satisfies the initial condition
\begin{equation}
P(x,0)=\delta(x-a)\,.
\label{eq:IPP2}
\end{equation}

\begin{figure}[H]
\begin{center}
\includegraphics{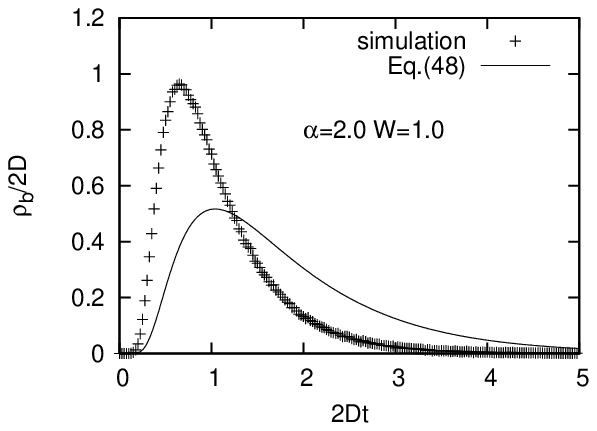}\includegraphics{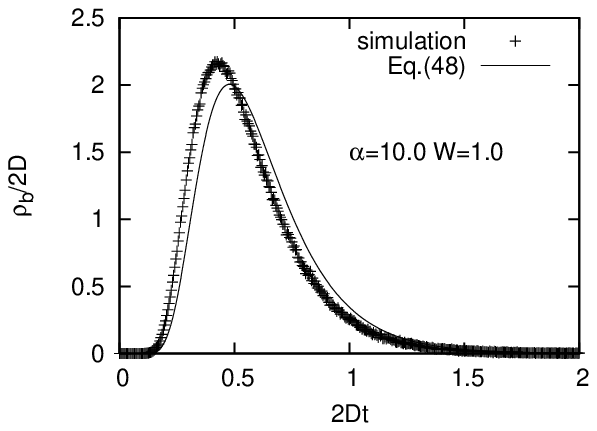}

\includegraphics{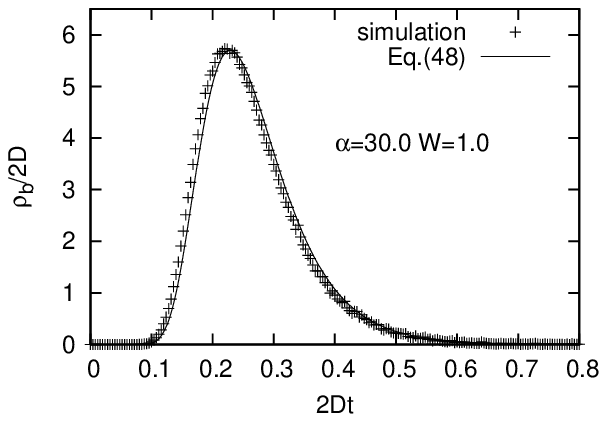}\includegraphics{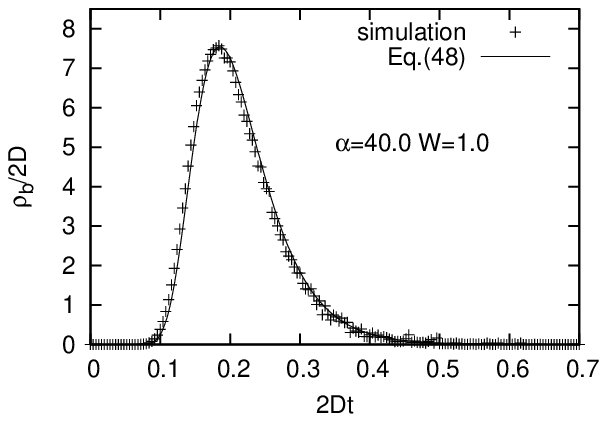}

\caption{\label{figSIPP}The event-duration distribution $\rho_{b}$ for inverted parabolic 
potentials with different dimensionless barrier heights $\frac{\alpha}{2}$. The absorbing walls are at
$-W$ and $W$, with $W=1$. The data are from path-sampling simulation and numerical
evaluations (lines) use the approximate formula, Eq.(\ref{eq:IPP5}). This approximation 
improves with increasing barrier height.}
\end{center}
\end{figure}

For the same potential with absorbing walls at $-W$ and $W$,
we approximate the current from  Eqs.(\ref{eq:IPP1}) and 
(\ref{eq:FPE3}) with $a=-W$ and $b=W$,
\begin{equation}
J(W,t)=\left(\frac{D\alpha W}{4}\sqrt{\frac{\alpha}{\pi}}\right)\frac{\exp\left[-\left(\alpha W^{2}/2\right)\coth\left(\alpha Dt/2\right)\right]}{\sinh\left(\alpha Dt/2\right)\sqrt{\sinh(\alpha Dt)}}\,.
\label{eq:IPP3}
\end{equation}
For normalization we will need 
\begin{equation}
N=\int_{0}^{\infty}J(W,t')dt'\,,
\label{eq:IPP3.1}
\end{equation}
which is the total probability passing to the right of $x=W$. Under the influence of this 
inverted parabolic potential, this probability will not pile up
but will flow toward $x\to\infty$. Thus
\begin{equation}
N=\Pi_{\infty}(-W)=\frac{\int_{-\infty}^{-W}\exp(-\frac{1}{2}\alpha x^{2})dx}{\int_{-\infty}^{\infty}\exp(-\frac{1}{2}\alpha x^{2})dx}=\frac{\mbox{erfc}(W\sqrt{\alpha/2})}{2}\,,
\label{eq:IPP4}
\end{equation}
from which one obtains the approximation
\begin{equation}
\rho_{b}(t)\simeq\left\{\frac{D\alpha W}{2[1-\mbox{erf}(W\sqrt{\alpha/2})]}\right\}\left(\sqrt{\frac{\alpha}{\pi}}\right)\frac{\exp\left[-\left(\alpha W^{2}/2\right)\coth\left(\alpha Dt/2\right)\right]}{\sinh\left(\alpha Dt/2\right)\sqrt{\sinh(\alpha Dt)}}\,,
\label{eq:IPP5}
\end{equation}
where $\mbox{erf}(x)=1-\mbox{erfc}(x)=\frac{2}{\sqrt{\pi}}\int_{0}^{x}\exp(-z^2)dz$
\cite{Erf-MathWorld}.
In Fig.~\ref{figSIPP} we compare the results from direct simulation
and from Eq.(\ref{eq:IPP5}) for inverted parabolic potentials
with different heights. In the simulations the two absorbing walls are placed at $a=-1.0$
and $b=1.0$; then the height of the barrier is given by  $\alpha/2$.

As expected, this approximation improves with increasing barrier height.

\subsection{Moments of the Distribution of Transition-Event Durations}

\subsubsection{Recursive Formula\label{sec:Rect_Formu}}
In studying a distribution, it is natural to investigate its moments.
Gardiner
~\cite{Gardiner-Stochastic_Methods_Handbook-E2nd-1994} 
provides an expression for the first moment of $\rho_{b}(t)$ (i.e., the mean time),
which is also given by Hummer
~\cite{Hummer-Transition_Path-JCP-2004}. 
Here we derive
a recursive formula for all \emph{moments} of $\rho_{b}(t)$, as suggested by Gardiner
~\cite{Gardiner-Stochastic_Methods_Handbook-E2nd-1994}.

Following Gardiner, we define $g_{b}(x,t)$ as the total probability that
the particle is absorbed at $b$ \textit{after} time $t$, given that it is released
at position $x$ at $t=0$. Thus
\begin{equation}
g_{b}(x,t) \equiv \int_{t}^{\infty}J(b,\tau)d\tau\,,
\label{eq:ReFo1}
\end{equation}
and we have the initial condition
\begin{equation}
P(x',0)=\delta(x'-x)\,.
\label{eq:ReFo-1}
\end{equation}
The limiting cases for $g_{b}(x,t)$ are
\begin{eqnarray}
g_{b}(x,t=0) & = & \Pi_{b}(x)\nonumber\\
g_{b}(x,t=\infty) & = & 0\,,
\label{eq:ReFo3}
\end{eqnarray}
where $\Pi_{b}(x)$ is defined in Eq.(\ref{eq:DTE8}) .

The $nth$ moment, $T_{n}$, of the exit time distribution for particles released at 
arbitrary $a<x<b$ can be calculated
from $g_{b}(x,t)$ according to
\begin{equation}
T_{n}(b,x)=-\int_{0}^{\infty}t^{n}\frac{\partial}{\partial t}\left[\frac{g_{b}(x,t)}{g_{b}(x,0)}\right]dt=\frac{n}{\Pi_{b}(x)}\int_{0}^{\infty}t^{n-1}g_{b}(x,t)dt\,,
\label{eq:ReFo4}
\end{equation}
so that $T_{n}(b,a)$ is the $nth$ moment of $\rho_{b}$. Gardiner 
~\cite{Gardiner-Stochastic_Methods_Handbook-E2nd-1994} 
shows that $g_{b}(x,t)$ satisfies the backward Fokker-Planck equation 
\begin{equation}
\left(-D\frac{dU^{\star}}{dx}\right)\frac{\partial g_{b}(x,t)}{\partial x}+D\frac{\partial^{2}g_{b}(x,t)}{\partial x^{2}}=\frac{\partial g_{b}(x,t)}{\partial t}\,.
\label{eq:ReFo5}
\end{equation}
Multiplying by $nt^{n-1}$ on both sides and integrating with respect
to $t$ yields
\begin{equation}
\left(-D\frac{dU^{\star}}{dx}\right)\left[n\int_{0}^{\infty}t^{n-1}\frac{\partial g_{b}(x,t)}{\partial x}dt\right]+D\left[n\int_{0}^{\infty}t^{n-1}\frac{\partial^{2}g_{b}(x,t)}{\partial x^{2}}dt\right]=n\int_{0}^{\infty}t^{n-1}\frac{\partial g_{b}(x,t)}{\partial t}dt\,.
\label{eq:ReFo6}
\end{equation}
Now the right side can be integrated by parts to find 
\begin{equation}
n\int_{0}^{\infty}t^{n-1}\frac{\partial g_{b}(x,t)}{\partial t}dt=-n\Pi_{b}(x)T_{n-1}(b,x)\,,
\label{eq:ReFo7}
\end{equation}
and with Eq.(\ref{eq:ReFo4}),
\begin{equation}
\left(-D\frac{dU^{\star}}{dx}\right)\frac{dy(x)}{dx}+D\frac{d^{2}y(x)}{dx^{2}}=-n\Pi_{b}(x)T_{n-1}(b,x)\,,
\label{eq:ReFo8}
\end{equation}
where 
\begin{equation}
y(x)\equiv\Pi_{b}(x)T_{n}(b,x)\,.
\label{eq:ReFo9}
\end{equation}
The boundary conditions on $y(x)$ are
\begin{equation}
y(b)=y(a)=0\,.
\label{eq:ReFo10}
\end{equation}

One way to solve equations like (\ref{eq:ReFo8}) uses Green's
functions
~\cite{Arfken_Weber-Mathematical_methods-E5th-2001}. 
The function that satisfies the homogeneous equation corresponding to
Eq.(\ref{eq:ReFo8}) with boundary condition (\ref{eq:ReFo10}) at $a$ is,
in fact, $\Pi_{a}(x)$; correspondingly at $b$, it is $\Pi_{b}(x)$.
Using these solutions, one obtains a recursive formula for all the moments
\begin{eqnarray}
T_{n}(b,x) & = & \frac{n}{D}\left\{ \int_{a}^{b}\exp[U^{\star}(x')]dx'\right\} \left\{ \frac{\Pi_{a}(x)}{\Pi_{b}(x)}\int_{a}^{x}\exp[-U^{\star}(x')]\Pi_{b}^{2}(x')T_{n-1}(b,x')dx'\right.\nonumber \\
 &  & \left.+\int_{x}^{b}\exp[-U^{\star}(x')]\Pi_{a}(x')\Pi_{b}(x')T_{n-1}(b,x')dx'\right\} \,.
\label{eq:ReFo11}
\end{eqnarray}
Our main interest is in the moments of $\rho_{b}$, namely,
\begin{eqnarray}
T_{n}(b,a) & = & \frac{n}{D}\left\{ \int_{a}^{b}\exp[U^{\star}(x)]dx\right\} \left\{ \int_{a}^{b}\exp[-U^{\star}(x)]\Pi_{a}(x)\Pi_{b}(x)T_{n-1}(b,x)dx\right\}\,.
\label{eq:ReFo12}
\end{eqnarray}
Given the moments according to  Eq.(\ref{eq:ReFo12}), the distribution
of transition-event durations, $\rho_{b}(t)$, can be reconstructed numerically,
at least for a fixed range of $t$.

\subsubsection{Lowest Eigenvalue\label{sec:first_eigen}}

From Eqs.(\ref{eq:FPE3}) and (\ref{eq:EM1}) one knows that $\rho_{b}(t)$
can be written in the series
\begin{equation}
\rho_{b}(t)=\sum_{n=1}^{\infty}C_{n}e^{-\lambda_{n}t}\,,
\label{eq:FEV1}
\end{equation}
where the eigenvalues defined by Eq.(\ref{eq:EM2}) satisfy $0<\lambda_{1}<\lambda_{2}<\lambda_{3}<...$.
The eigenvalues, in particular $\lambda_1$, can be found via direct numerical solution of 
Eq.(\ref{eq:EM2}). Here we show an alternative based on integrations involving the potential.
The first eigenvalue $\lambda_{1}$ can be expressed in terms
of the high-order moments because of asymptotically exponential behavior. 
When $n\gg1$,
\begin{eqnarray}
T_{n}(b,a) & = & \int_{0}^{\infty}t^{n}\rho_{b}(t)dt\nonumber \\
 & = & \Gamma(n+1)\frac{C_{1}}{\lambda_{1}^{n+1}}\left[1+\frac{C_2}{C_1}\left(\frac{\lambda_1}{\lambda_2}\right)^{n+1}+\frac{C_3}{C_1}\left(\frac{\lambda_1}{\lambda_3}\right)^{n+1}+...\right]\,,
\label{eq:FEV2}
\end{eqnarray}
where $\Gamma(n)$ is the Gamma function. The lowest eigenvalue 
can then be estimated from a ratio of high moments; for example,
\begin{equation}
\lambda_{1}=\lim_{n\rightarrow\infty}\left[\frac{T_{n}(b,a)}{T_{n+1}(b,a)}(n+1)\right]\,,
\label{eq:FEV3}
\end{equation}
and from Eq.(\ref{eq:FEV2}), the constant can be determined according to
\begin{equation}
C_1=\lim_{n\rightarrow\infty}\frac{T_n(b,a)\lambda_1^{n+1}}{\Gamma(n+1)}\,.
\label{eq:FEV3.1}
\end{equation}
Recalling that the moments can be constructed via successive integration, Eq.(\ref{eq:FEV3}) 
provides a way to estimate the first eigenvalue in 
Eq.(\ref{eq:FEV1}). 
In Section~\ref{sec:long_time}
Eq.(\ref{eq:FEV3}) will be used together with simulations to 
check an approximate analytic result 
for the leading eigenvalue in a representative case.

\subsubsection{First Moment}

For $n=1$, using $T_{0}(b,a)=1$, Eq.(\ref{eq:ReFo12}) yields
the first moment of the distribution of transition-event durations,
\begin{equation}
T_{1}(b,a)=\frac{1}{D}\left\{ \int_{a}^{b}\exp[U^{\star}(x)]dx\right\} \left\{ \int_{a}^{b}\exp[-U^{\star}(x)]\Pi_{a}(x)\Pi_{b}(x)dx\right\}\,.
\label{eq:FM1}
\end{equation}
We can immediately evaluate $T_1$ for the simple potentials.
For free diffusion with $U^\star=0$ and $a=0$, $b=L$, 
\begin{equation}
\frac{2DT_1^0}{L^2}=\frac{1}{3}\,.
\label{eq:FM2}
\end{equation}
With $U^\star=kx$, $a=0$, $b=L$,
\begin{equation}
\frac{2DT_1}{L^2}=\frac{2}{(kL)^2}\left[kL\coth\left(\frac{kL}{2}\right)-2\right]\,.
\label{eq:FM2.1}
\end{equation}

For an inverted parabolic potential $U^\star=H(1-\frac{x^2}{W^2})$, where the curvature
$\alpha=\frac{2H}{W^2}$, and $a=-W$, $b=W$, we can find an approximation of $T_1$. When 
$H \gg 1$, by using the method of steepest descents,
\begin{equation}
\int_{-W}^{W}\exp[U^{\star}(x)]dx \approx \exp(H)\sqrt{\frac{2\pi}{\alpha}}\,,
\label{eq:FM3}
\end{equation}
and
\begin{equation}
\Pi_a(x)\Pi_b(x) \approx \frac{1-\left[\mbox{erf}\left(\frac{x\sqrt{\alpha}}{\sqrt{2}}\right)\right]^2}{4}\,.
\label{eq:FM4}
\end{equation}
Then
\begin{eqnarray}
2DT_1 &\approx& \sqrt{\frac{2\pi}{\alpha}}\int_0^W\exp\left(\frac{Hx^2}{W^2}\right)\left\{1-\left[\mbox{erf}\left(\frac{\sqrt{H}x}{W}\right)\right]^2\right\}dx \nonumber\\
& = & \frac{\sqrt{\pi}W^2}{H}\int_0^{\sqrt{H}}\exp(y^2)\{1-[\mbox{erf}(y)]^2\}dy \nonumber\\
& = & \frac{\sqrt{\pi}W^2}{H}\left\{\int_0^{1}\exp(y^2)\{1-[\mbox{erf}(y)]^2\}dy+\int_1^{\sqrt{H}}\exp(y^2)\{1-[\mbox{erf}(y)]^2\}dy\right\} \nonumber\\
&\approx& \frac{W^2}{H}[1.27+\log(H)]=\frac{2}{\alpha}[1.27+\log(H)]\,.
\label{eq:FM5}
\end{eqnarray}
Eq.(\ref{eq:FM5}) provides a good approximation for $T_1$ for the inverted parabolic potential with 
high barrier. It also gives a rough estimate of $T_1$ for a ``single-bump'' barrier with 
height $H$ and width $2W$. 

Naively, one might guess $T_1$ should simply be proportional to the effective frequency for the
barrier top, namely, the inverse curvature $\alpha^{-1}$. However, this intuition falls short in 
two respects. First, the logarithmic term in Eq.(\ref{eq:FM5}) is dominant for large barriers, even
for this simplest purely parabolic potential. Further, in extensive numerical work, for a double-well
potential, we have seen unambiguously that the mean event duration is sensitive to details of the 
potential far from the barrier top (data not shown). This sensitivity can be traced to the dependence
of the optimal ``speed'', Eq.(\ref{eq:PI10}), on details of the potential. To give an extreme 
example, if there were a second barrier and minimum in the potential, then $T_1$ would have to 
include the Kramers' time for the second barrier.

\subsubsection{Reconstruction of $\rho_b$ from Moments\label{sec:Res_Den}}
Reconstructing a function approximately from a finite number of moments has been 
studied, e.g., by maximum entropy method
~\cite{Mead_Papanicolaou-Maximum_Entropy-JMP-1984,
Tagliani-Maximum_Entropy-JMP-1993,
Poland-moments-JCP-1995},
continued fraction approach
~\cite{Jhon_Dahler-Moments_Continued_Fraction-JCP-1978,
Macchi_Tognetti-Moments_Continued_Fraction-PRB-1996},
and Talenti method
~\cite{Talenti-moments_2_function-IP-1987,
Hon-Wei_Gram-Schmidt-moments_EABE_2002} 
and perhaps other techniques. Here we follow Hon and Wei's work
~\cite{Hon-Wei_Gram-Schmidt-moments_EABE_2002} 
to reconstruct the density $\rho_b(t)$ in a similar way.

First one builds up an orthonormal set of basis functions $\psi_j(t)$, which are polynomials, 
\begin{equation}
\psi_j(t)=\sum_{n=0}^{j}c_{jn}t^n\,,
\label{eq:RDM1}
\end{equation}
by using the standard Gram-Schmidt orthonormalization technique
~\cite{Arfken_Weber-Mathematical_methods-E5th-2001}. 
The polynomials satisfy
\begin{equation}
\int_0^\infty\psi_{j'}(t)\psi_j(t)w(t)dt=\delta_{jj'}\,,
\label{eq:RDM2}
\end{equation}
with respect to a weight function $w(t)$, which is tailored to our problem with the choice
\begin{equation}
w(t)=\exp(-\lambda_{1}t)\exp\left(-\frac{L^2}{4Dt}\right)(1+t^{-\frac{5}{2}})\,,
\label{eq:RDM3}
\end{equation}
where $L=b-a$.
For this weight factor, when $t\to \infty$, $w(t)\sim \exp(-\lambda_{1}t)$, and when $t\to 0$, 
$w(t)\sim \exp\left(-\frac{L^2}{4Dt}\right)t^{-\frac{5}{2}}$. These forms represent the long-time 
and short-time behaviors for $\rho_b(t)$ as we will show in Section~\ref{sec:long_time}
and (\ref{sec:free_diff}) below. Thus we 
built in all the information we know about $\rho_b(t)$ in this weight factor. Notice that $\lambda_{1}$ 
can be found by following the scheme in Section~\ref{sec:first_eigen}, or via other numerical methods.

Following the usual Gram-Schmidt procedure, one builds up 
\begin{equation}
\rho_m(t)=w(t)\sum_{j=0}^ma_j\psi_j(t)\,,
\label{eq:RDM4}
\end{equation}
which will be an approximation for $\rho_b(t)$ of ``order'' $m$. If we incorporate moments of 
$\rho_b$ by setting coefficients according to
\begin{eqnarray}
a_0 & = & \frac{T_0}{\int_0^\infty \psi_0(t)w(t)dt}=\frac{1}{c_{00}\int_0^\infty w(t)dt} \nonumber \\
a_j & = & \frac{T_j-\sum_{i=0}^{j-1}a_{i}\int_0^\infty t^{j}\psi_{i}(t)w(t)dt}{\int t^{j}\psi_j(t)w(t)dt}\,,
\label{eq:RDM6}
\end{eqnarray}
then $\rho_m(t)$ will reproduce the first $m$ moments $T_n$ ($0\le n \le m$), i.e.,
\begin{equation}
T_n \equiv \int_{0}^{\infty}t^n\rho_b(t)dt \approx \int_{0}^{\infty}t^n\rho_m(t)dt\,.
\label{eq:RDM5}
\end{equation}

\begin{figure}[H]
\begin{center}

\includegraphics{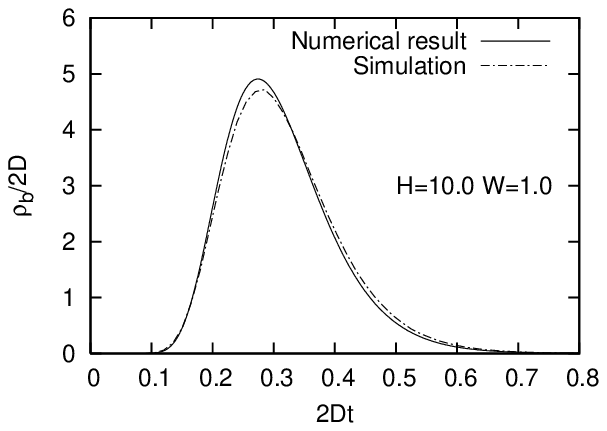}\includegraphics{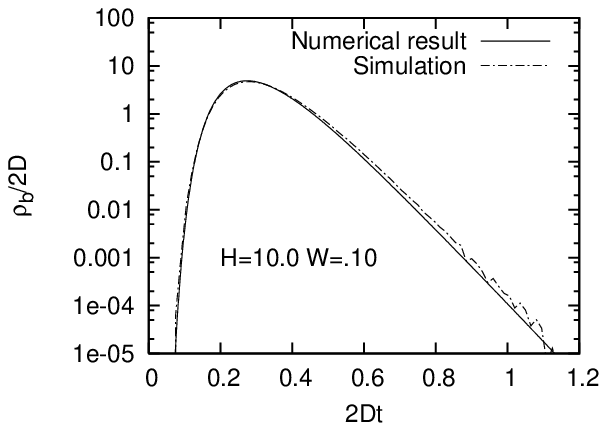}

\includegraphics{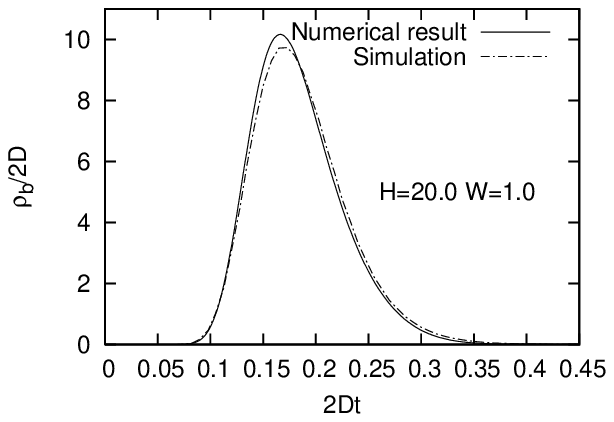}\includegraphics{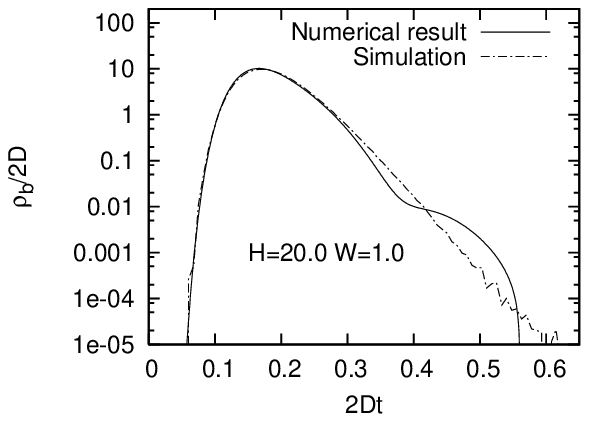}

\caption{\label{figM2D}Reconstruction of $\rho_b$ for double-well potentials 
$U^{\star}=H\left[\left(\frac{x}{W}\right)^2-1\right]^2$ based on moments and a 
Gram-Schmidt procedure. The two absorbing walls
are at $-W$ and $W$, where $W=1$. The data are from path-sampling simulations and numerical
results are based on theory in Section~\ref{sec:Rect_Formu} and \ref{sec:Res_Den},
using the first five moments. They match well
except the long time tail as emphasized in the semi-log plot.}
\end{center}
\end{figure}

By using the first five moments, we reconstruct the distribution of transition-event 
durations for several different 
double-well potentials. Two of the results are shown in Fig.~\ref{figM2D}. 
They match well with the simulations
except for the long time tail as seen in the semi-log plot. However, the event 
probability in that region is quite small.

\subsection{Short Time Behavior}
Beyond the moments, it is of interest to study the asymptotic behavior of the event-duration
distribution, in both short and long time limits. We first analyze the $t\to 0$ behavior, 
using exact methods.

\subsubsection{Short Time Behavior for the Green Function with Open Boundary Conditions}

From the perspective of the path integral, introduced in Section~\ref{sec:Path_Integral}, 
if $(t-t_{0})$ is short, the velocity on the optimal trajectory $\frac{dx_{c}}{dt}$ 
will be large. This implies a large ``energy'' $E$ in Eq.(\ref{eq:PI8}).
We therefore assume $2E\gg4DV$ and obtain
the corresponding Green function for short time $t$ with open boundary conditions,
\begin{eqnarray}
G(x,t|a,t_{0}) & \approx & \exp\left(\frac{1}{2k_{B}T}\int_{a}^{x}Fdx'\right)\frac{1}{\sqrt{4\pi D(t-t_{0})}}\exp\left[-\frac{(x-a)^{2}}{4D(t-t_{0})}\right]\nonumber \\
& \times &\exp[-\overline{V}(t-t_{0})]\,,
\label{eq:STBO1}
\end{eqnarray}
where $\overline{V}$ is the average effective potential between $a$
and $x$,
\begin{equation}
\overline{V}=\frac{1}{x-a}\int_{a}^{x}V(x')dx'\,.
\label{eq:STBO2}
\end{equation}
We have used the quadratic approximation, which is expected to be reasonable if 
the diffusion coefficient $D$ is small
~\cite{Wiegel-Path_Integral-1986}.
Also, because we are further restricting our analysis to short-time behavior, 
the important paths will be close to the optimal one, which should improve the approximation.

If we define $G_{0}$ as the Green function for free diffusion with open boundary
conditions
~\cite{Gardiner-Stochastic_Methods_Handbook-E2nd-1994},
\begin{equation}
G_{0}(x,t|a,t_{0})=\frac{1}{\sqrt{4\pi D(t-t_{0})}}\exp\left[-\frac{(x-a)^{2}}{4D(t-t_{0})}\right]\,,
\label{eq:STBO3}
\end{equation}
Eq.(\ref{eq:STBO1}) can be expressed as
\begin{equation}
G(x,t|a,t_{0}) \approx G_{0}(x,t|a,t_{0})\exp\left(\frac{1}{2k_{B}T}\int_{a}^{x}Fdx'\right)\exp[-\overline{V}(t-t_{0})]\,,
\label{eq:STBO4}
\end{equation}
which will be useful when we discuss the early time behavior with alternative 
boundary conditions below in Section~\ref{sec:Short_Time_Spetial}.

\subsubsection{Short Time Behavior for the Current with Absorbing Boundary Conditions\label{sec:Short_Time_Spetial}}

In this subsection we need to retrace the path integral method in order to study 
transition-event durations, as required, with two absorbing walls.

\begin{figure}[H]
\begin{center}
\includegraphics{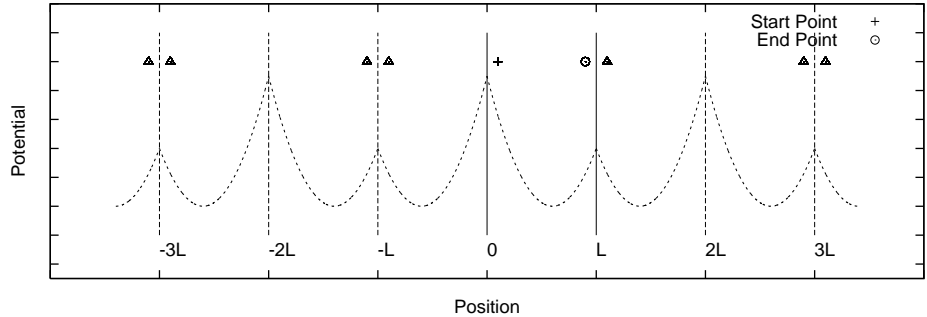}

\caption{\label{figSTBS1}Calculating the path integral between the
start point '$+$' and the end point '$\circ$' with absorbing walls at $x=0$
and $x=L$. The dashed curve represents an arbitrary potential between the two absorbing walls.
Eq.(\ref{eq:STBS1}) indicates that one must calculate the path integrals between 
the start point '$+$' and all ``end points'', including the real one '$\circ$' and
image end points '$\triangle$', in the periodic potential with open 
boundary condition.}
\end{center}
\end{figure}

As shown in Fig.~\ref{figSTBS1}, we wish to calculate the path integral from
the start point '$+$' at $x=0+$ to the end point '$\circ$' at $x=L-$
during the time interval ($t-t_{0}$). 
There are two absorbing walls at position $x=0$ and $x=L$, with some arbitrary
potential between them. If the position of the start point is $\epsilon_1$
and the end point is $x$, 
then the Green function $G^{abs}$ with absorbing boundary conditions
can be expressed as a sum over Green functions $G$ for open boundary conditions
~\cite{Kleinert-Path_Integral-2004}
\begin{equation}
G^{abs}(x,t|\epsilon_1,t_{0})=\sum_{j=-\infty}^{\infty}G(2jL+x,t|\epsilon_{1},t_{0})-\sum_{j=-\infty}^{\infty}G(2jL-x,t|\epsilon_{1},t_{0})\,.
\label{eq:STBS1}
\end{equation}
The construction is shown schematically in Fig.~\ref{figSTBS1}.

As shown above, to determine the distribution of event durations, we first calculate the
current at $x$, then take the limits $\epsilon_{1}\rightarrow0$ and $x\rightarrow L$,
\begin{equation}
J^{abs}(L,t|0,t_{0})=-2D\sum_{j=-\infty}^{\infty}\left.\frac{\partial G(x,t|0,t_{0})}{\partial x}\right|_{x=(2j+1)L}\,.
\label{eq:STBS4}
\end{equation}
From Eq.(\ref{eq:STBO4}), when $(t-t_{0})$ is small
\begin{eqnarray}
\left.\frac{\partial G(x,t|0,t_{0})}{\partial x}\right|_{x=(2j+1)L} & \approx & \left.\frac{\partial G_{0}(x,t|0+,t_{0})}{\partial x}\right|_{x=(2j+1)L}\times\exp\left[\frac{1}{2k_{B}T}\int_{0}^{(2j+1)L}Fdx'\right]\nonumber \\
 &  & \times\exp[-\overline{V_{j}}(t-t_{0})]\,,
\label{eq:STBS5}
\end{eqnarray}
where the symmetry of the periodically continued potential has been used; see Fig.~\ref{figSTBS1}.

The periodicity implies further simplifications, including
\begin{eqnarray}
\int_{0}^{(2j+1)L}Fdx'&=&\int_{0}^{L}Fdx'\nonumber \\
\overline{V_{j}}=\frac{1}{(2j+1)L}\int_{0}^{(2j+1)L}V(x')dx'&=&\frac{1}{L}\int_{0}^{L}V(x')dx'=\overline{V_{0}}\,.
\label{eq:STBS7}
\end{eqnarray}
If we define the free diffusion current with absorbing boundary conditions as
\begin{equation}
J^{abs}_{0}(L,t|0,t_{0})=-2D\sum_{j=-\infty}^{\infty}\left.\frac{\partial G_{0}(x,t|0,t)}{\partial x}\right|_{x=(2j+1)L}\,,
\label{eq:STBS8}
\end{equation}
Eq.(\ref{eq:STBS4}) can be written
\begin{equation}
J^{abs}(L,t|0,t_{0}) \approx J^{abs}_{0}(L,t|0,t_{0})\exp\left(\frac{1}{2k_{B}T}\int_{0}^{L}Fdx'\right)\exp[-\overline{V_{0}}(t-t_{0})]\,,
\label{eq:STBS9}
\end{equation}
or
\begin{equation}
J^{abs}(b,t|a,t_{0}) \approx J^{abs}_{0}(b,t|a,t_{0})\exp\left(\frac{1}{2k_{B}T}\int_{a}^{b}Fdx'\right)\exp[-\overline{V}(t-t_{0})]
\label{eq:STBS10}
\end{equation}
where
\begin{equation}
\overline{V}=\frac{1}{b-a}\int_{a}^{b}V(x')dx'\,.
\label{eq:STBS11}
\end{equation}
We can now estimate the short time behavior of the normalized current.

\subsubsection{Short Time behavior for the distribution of the transition-event durations}

Combining Eq.(\ref{eq:DTE7}) and (\ref{eq:STBS10}) , we find that the
short-time behavior for the distribution of transition-event durations is given by
\begin{equation}
\rho_{b}(t\to 0) \approx \rho_{b}^{(0)}(t)\exp\left(\frac{1}{2k_{B}T}\int_{a}^{b}Fdx'\right)\exp(-\overline{V}t)\left[\lim_{\epsilon\rightarrow0+}\frac{\Pi_{b}^{0}(a+\epsilon)}{\Pi_{b}(a+\epsilon)}\right]\,,
\label{eq:STBDT1}
\end{equation}
where $\Pi_{b}^{0}(x)$ is the splitting probability for free diffusion and 
$\rho_{b}^{(0)}$ is the distribution for free-diffusion with absorbing
boundary conditions. From Eq.(\ref{eq:DTE8})
\begin{eqnarray}
\lim_{\epsilon\rightarrow0+}\frac{\Pi_{b}^{0}(a+\epsilon)}{\Pi_{b}(a+\epsilon)}=\frac{\int_{a}^{b}\exp[U^{\star}(x')]dx'}{\exp[U^{\star}(a)](b-a)}\,.
\label{eq:STBDT2}\end{eqnarray}
Combining Eq.(\ref{eq:STBDT1}) and (\ref{eq:STBDT2}), and using
$F(x)=k_{B}T\left(-\frac{dU^{\star}}{dx}\right)$, the normalized distribution becomes
\begin{equation}
\rho_{b}(t\to 0) \approx \rho_{b}^{(0)}(t)\left\{ \frac{\int_{a}^{b}\exp[U^{\star}(x')]dx'}{\exp[U^{\star}(a)](b-a)}\right\} \exp\left[\frac{U^{\star}(b)-U^{\star}(a)}{2}\right]\exp(-\overline{V}t)\,,
\label{eq:STBDT3}
\end{equation}
revealing corrections to the free diffusion result due to the potential.

\begin{figure}[H]
\begin{center}

\includegraphics{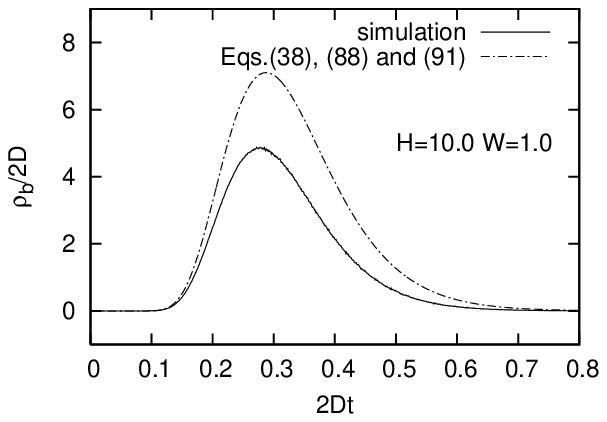}\includegraphics{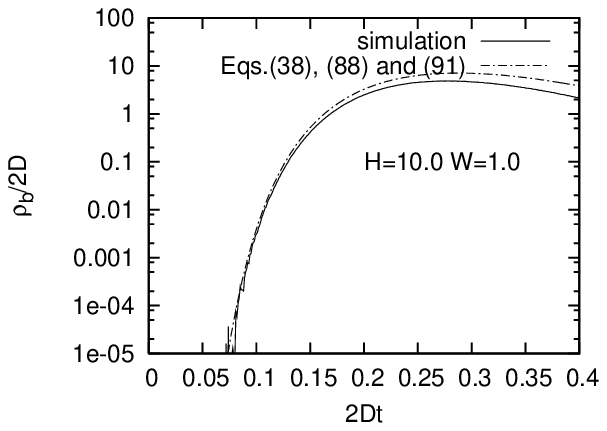}

\includegraphics{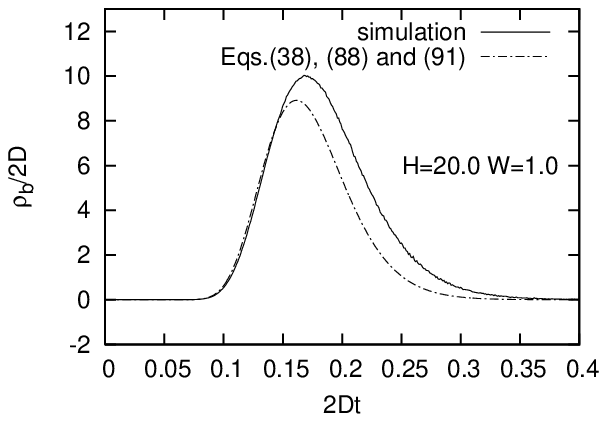}\includegraphics{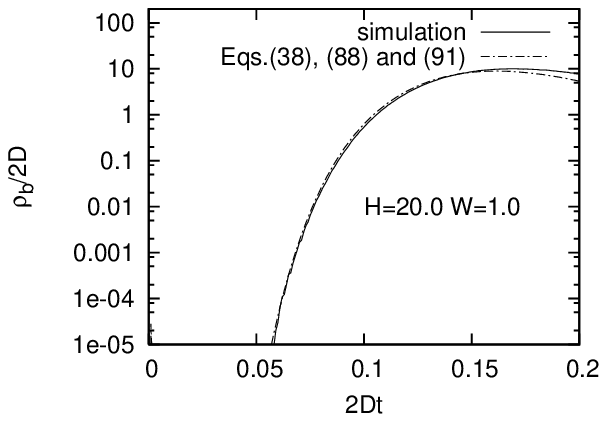}

\caption{\label{figSSB}Short time behavior of $\rho_b$ for double-well potentials 
$U^{\star}=H\left[\left(\frac{x}{W}\right)^2-1\right]$ with two absorbing walls
at $-W$ and $W$, where $W=1$. Data from path-sampling simulations (solid lines) are compared
to numerical estimation of $\rho_{b}(t\to 0)$ (dashed lines) based on
Eqs.(\ref{eq:FU1}), (\ref{eq:STBS11}) and (\ref{eq:STBDT3}) to
numerically estimate the short time behavior of $\rho_b$. }
\end{center}
\end{figure}

In Fig.~\ref{figSSB} we compare the results from a path-sampling simulation as described 
in Sec.~\ref{sec:path-sampling} 
and our final result Eq.(\ref{eq:STBDT3}) for two double-well potentials of varying 
barrier height. The simulations and the analytic results of Eq.(\ref{eq:STBDT3}) are in good agreement at
sufficiently early times, although at the earliest times the simulations reveal degrading statistics.

\subsubsection{Comment on the Short Time Behavior of $\rho_{b}^{(0)}(t)$\label{sec:free_diff}}

For the free diffusion problem with absorbing walls, the path integral method 
in Section~\ref{sec:Short_Time_Spetial} can give 
the solution. Redner~\cite{Redner-1st_Passage-2001} shows that an image
method, similar to that used in electromagnetism, will give the same
result

\begin{equation}
\rho_{b}^{(0)}(t)=\frac{2L}{\sqrt{\pi Dt^3}}\sum_{j=0}^{\infty}\left[\frac{(2j+1)^2L^2-2Dt}{4Dt}\right]\exp\left[-\frac{(2j+1)^2L^2}{4Dt}\right]\,.
\label{eq:STBF6}
\end{equation}
If $Dt \ll L^2$,
\begin{equation}
\rho_{b}^{0}(t)\sim t^{-\frac{5}{2}}\exp\left(-\frac{L^{2}}{4Dt}\right)\,.
\label{eq:STBF7}
\end{equation}

Eq.(\ref{eq:STBF7}) shows the short time behavior of $\rho_{b}$ for 
free diffusion with absorbing boundary conditions. Combining it with Eq.(\ref{eq:STBDT3}), 
one can find the short time behavior in the presence of the potential.

\subsection{Long Time Behavior for Double-Well Potential\label{sec:long_time}}

From Eqs.(\ref{eq:FPE3}) and (\ref{eq:EM1}) we know that the long
time behavior of $\rho_{b}(t)$ will be determined primarily by the first
eigenvalue $\lambda_{1}$. Here we use a perturbative approach to
obtain an approximation for $\lambda_{1}$ for a double-well potential with high
barrier.
We confirm, by direct numerical calculation, the validity of using perturbation 
theory. We also perform path-sampling simulations to check the accuracy of our 
final approximation.

By a variable transformation, the one dimensional Fokker-Planck equation
can be transformed to a Schr\"{o}dinger-like equation
~\cite{Risken-Fokker_Planck_Equation-E2nd-1989}. 
If we let $\phi_n(x)=\exp(\frac{U^{\star}}{2})p_n(x)$,
Eq.(\ref{eq:EM2}) becomes
\begin{equation}
\left[-D\frac{d^2}{dx^2}+V_{s}(x)\right]\phi_n(x)=\lambda_n \phi_n(x)\,,
\label{eq:LTB0.1}
\end{equation}
where
\begin{equation}
V_{s}(x)=\frac{D}{4}\left(\frac{dU^{\star}(x)}{dx}\right)^{2}-\frac{D}{2}\frac{d^{2}U^{\star}(x)}{dx^{2}}\,,
\label{eq:LTB1}
\end{equation}
which is exactly the effective potential in Eq.(\ref{eq:PI9}).
The eigenvalue spectrum remains the same
~\cite{Risken-Fokker_Planck_Equation-E2nd-1989}. 

We write a double-well potential in the form
\begin{equation}
U^\star=H\left[1-\left(\frac{x}{W}\right)^2\right]^2\,,
\label{eq:LTB1.1}
\end{equation}
where $H$ is the height of the barrier in units of $k_B T$ and $W$ is the half-width of the 
barrier. We will consider
double-wells having high barriers, $H \gg 1$, and fixed half width, $W$.
The absorbing walls are placed at the two minima, $x=\pm W$. The Schr\"{o}dinger 
potential corresponding to (\ref{eq:LTB1.1}) is 
\begin{equation}
V_{s}(x)=D\left[\frac{2H}{W^2}+\left(\frac{4H^2}{W^4}-\frac{6H}{W^4}\right)x^{2}-\frac{8H^2}{W^6}x^{4}+\frac{4H^2}{W^8}x^{6}\right]\,.
\label{eq:LTB2}
\end{equation}

\begin{figure}[H]
\begin{center}
\includegraphics{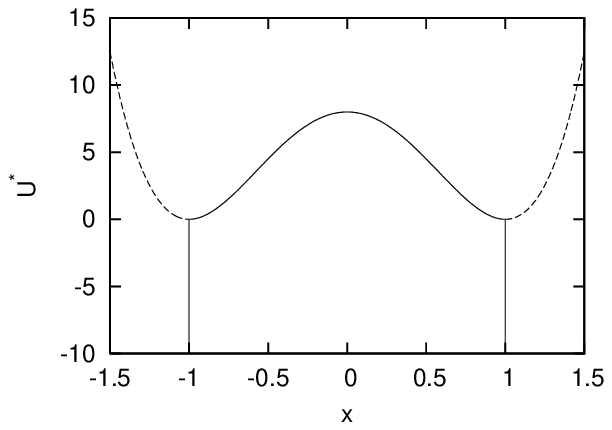}\includegraphics{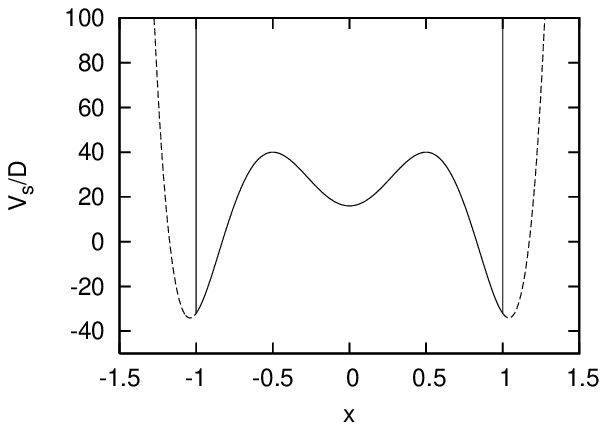}

\caption{\label{figLTB1}A high barrier double-well potential (left) and its Schr\"{o}dinger
analogue (right) from Eqs.(\ref{eq:LTB1.1}) and (\ref{eq:LTB2}), for $H=8$ and $W=1$.
The two absorbing walls are put at the two minima. 
If the barrier height of the potential is not sufficiently
large, the minimum of $V_s$ at $x=0$ will disappear.}
\end{center}
\end{figure}

In Fig.\ref{figLTB1} we plot $U^{\star}(x)$ and $V_{s}(x)$ for a
double-well with sufficiently high barrier.
For the potential in the Schr\"{o}dinger picture, $V_s=+\infty$ outside the central 
interval to ensure the wave functions $\phi_n(x)$ vanish at the ends of the interval,
thus satisfying the absorbing boundary conditions.

We use perturbation theory to describe the lowest stationary state, which must exist because of 
the boundary conditions. For sufficiently high barriers, we expect that, the lowest eigenstate will be 
localized at the central minimum, suggesting the use of a perturbation process based on a 
simple harmonic oscillator. Using a numerical procedure for bound-state solutions of the 
time-independent Schr\"{o}dinger equation
~\cite{Gould_Tobochnik-Introduction_Computer_Simulation_Methods-E2nd-1995}, 
we confirmed this localization 
for high barriers. We also note that for high barriers, the oscillator's 
Gaussian wave function nearly vanishes
at the boundaries, in approximate satisfaction of the proper boundary conditions.

Our perturbation calculation is therefore based on separating off the dominant harmonic component 
of $V_s$, noting $H \gg 1$, using
\begin{equation}
V_{s}(x)=V_{0}(x)+V_{1}(x)\,,
\label{eq:LTB3}
\end{equation}
where 
\begin{eqnarray}
V_{0}(x) & = & D\left(\frac{2H}{W^2}+\frac{4H^2}{W^4}x^{2}\right)\,,
\label{eq:LTB4} \\
V_{1}(x) & = & D\left(-\frac{6H}{W^4}x^{2}-\frac{8H^2}{W^6}x^{4}+\frac{4H^2}{W^8}x^{6}\right)\,.
\label{eq:LTB5}
\end{eqnarray}
From textbook results for a linear harmonic oscillator, the first eigenvalue and wave 
function are
\begin{eqnarray}
\lambda_{1}^{(0)} & = & D\left(\frac{4H}{W^2}\right)\,,
\label{eq:LTB6} \\
\psi_{1}^{(0)}(x) & = & \frac{\sqrt{\gamma}}{\pi^{1/4}}\exp\left(-\frac{1}{2}\gamma^{2}x^{2}\right)\,,
\label{eq:LTB7}
\end{eqnarray}
where 
\begin{equation}
\gamma=\left(\frac{2H}{W^2}\right)^{1/2}\,.
\label{eq:LTB8}
\end{equation}
The first order perturbative correction is
\begin{equation}
\lambda_{1}^{(1)}=\int\psi_{1}^{(0)*}V_{1}(x)\psi_{1}^{(0)}dx=D\left[-\frac{3}{W^2}+\mathcal{O}\left(\frac{1}{HW^2}\right)\right]\,,
\label{eq:LTB13}
\end{equation}
which is down by a factor of $H$ from the zero order result.
The second order correction is down by another factor of $H$:
\begin{equation}
\lambda_{1}^{(2)}={\sum_{m}}'\left\{\frac{|V_{m1}|^{2}}{\lambda_{1}^{(0)}-\lambda_{m}^{(0)}}\right\}\sim D\left[\mathcal{O}\left(\frac{1}{HW^2}\right)\right]\,.
\label{eq:LTB14}
\end{equation}
The net result for the lowest eigenvalue is thus
\begin{eqnarray}
\lambda_{1} & = & D\left[\frac{4H}{W^2}-\frac{3}{W^2}+\mathcal{O}\left(\frac{1}{HW^2}\right)\right] 
\label{eq:LTB14.5}\\
& =& D\left[\alpha-\frac{3}{W^2}+\mathcal{O}\left(\frac{1}{HW^2}\right)\right]\,,
\label{eq:LTB15}
\end{eqnarray}
where $\alpha=4H/W^2$ gives the curvature at the top of the barrier.

Eq.(\ref{eq:LTB15}) shows that the long time behavior is simply linear in 
the barrier-top curvature $\alpha$, for large values of $\alpha$ and fixed $W$. This is also 
the case for the inverted parabolic 
potential, as can be determined from Eq.(\ref{eq:IPP5}), or by performing the same 
calculation as we did for the double-well in this section. In fact, when $Dt\alpha \gg 1$, 
Eq.(\ref{eq:IPP5}) becomes
\begin{equation}
\rho_{b}(t)\simeq\left[\frac{D\alpha W}{1-\mbox{erf}(W\sqrt{\alpha/2})}\right]\left(\sqrt{\frac{2\alpha}{\pi}}\right)\exp\left(-\frac{\alpha W^{2}}{2}\right)\exp(-D\alpha t)\,.
\label{eq:LTB15.5}
\end{equation}
We therefore expect similar
linearity with $\alpha$ in the higher barrier limit 
for any system that can be approximated by an inverted parabola 
and a high order correction.

\begin{figure}[H]
\begin{center}
\includegraphics{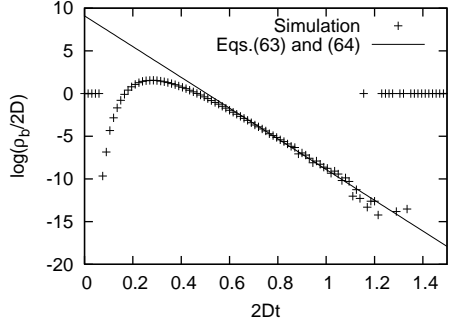}
\caption{\label{figSME}Comparison of two numerical approaches for computing the long
time behavior of the event-duration distribution. A path-sampling simulation is compared to
pure exponential behavior based on the lowest eigenvalue $\lambda_1$ determined from Eqs.(\ref{eq:FEV3})
and (\ref{eq:FEV3.1}) with $H=10$ and $W=1$. Similar results are obtained for other
high barrier cases.}
\end{center}
\end{figure}

We performed numerical checks of the approximation (\ref{eq:LTB15}). 
We determined the lowest eigenvalue $\lambda_{1}$ numerically using 
high moments and Eq.(\ref{eq:FEV3}). We also used path-sampling simulations as a consistency check.
Fig.~\ref{figSME} compares the numerical evaluation
of $\lambda_1$ via Eqs.(\ref{eq:FEV3}) and (\ref{eq:FEV3.1}) with a path-sampling simulation data 
for a particular double-well potential.  

\begin{figure}[H]
\begin{center}
\includegraphics{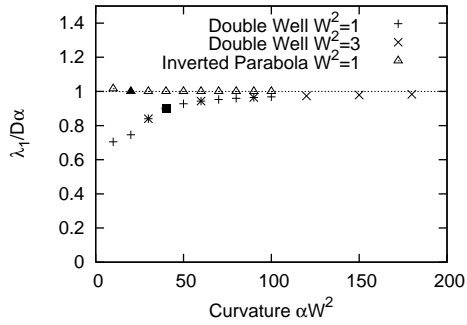}
\caption{\label{figLTS}Long time behavior of $\rho_b$ for double-well potentials and the inverted parabola.
Exact numerical results for lowest eigenvalue $\lambda_1/D\alpha$ are plotted as a function 
of the dimensionless curvature $\alpha W^2$ at the barrier peaks. 
Note that barrier heights are proportional to $\alpha$ for the models considered.
Two double-well potentials with fixed $W^2=1$ and $W^2=3$,
and one inverted parabola system with fixed $W^2=1$ 
are considered. Filled symbols indicate the $\alpha$ values where the barrier height $H$ is 
10 (in units of $k_BT$). The values of $\lambda_1/D\alpha$, 
which dominate the long-time behavior, approach $1$ 
for large curvatures, as predicted by Eq.(\ref{eq:LTB15}).}
\end{center}
\end{figure}

In Fig.~\ref{figLTS}, we compare Eq.(\ref{eq:LTB15}) with numerical calculations of $\lambda_1$ 
for double-well potentials and inverted parabolae  with fixed $W$ and a 
range of curvatures ($10 - 100$) and plot $\lambda_1/D\alpha$ as a function
of curvature $\alpha W^2$. As Eq.(\ref{eq:LTB15}) predicts, $\lambda_1/D\alpha$
approaches unity for large curvature.

\section{SUMMARY}

We have applied a combination of analytic and numerical techniques to study
the distribution, $\rho_{b}(t)$, of the durations of transition-events over a barrier in a 
one-dimensional system undergoing over-damped Langevin dynamics. The typical event duration is much 
shorter than the well-studied first passage time (FPT)
~\cite{Berezhkovskii_Hummer_Bezrukov-Equivalence_Trans_Paths_Ion_Channels-PRL-2006}; 
see Figs.~\ref{figINT1} and \ref{figINT2}. 
The event duration scale is the simplest non-trivial measure
of the detailed dynamics of an activated process, and we believe it is critical for future quantitative
study of dynamics of many chemical and biological systems
~\cite{Zuckerman-Butane-JCP-2002}.

The distribution $\rho_{b}(t)$ can be derived from 
the Fokker-Planck equation with special boundary conditions and was subjected to detailed
analysis. A number of results are obtained,
including:  
(i) the analytic form of the asymptotic short-time behavior ($t \rightarrow 0$), which is
universal and independent of the potential function;
(ii) the first non-universal correction to the short-time behavior; 
(iii) following Gardiner 
\cite{Gardiner-Stochastic_Methods_Handbook-E2nd-1994}, a recursive formulation 
for calculating, exactly, all moments of $\rho_{b}$ based solely on the potential function
--- along with approximations for the distribution based on a small number of moments; 
(iv) a high-barrier approximation to the long-time ($t \rightarrow \infty$) behavior of $\rho_{b}(t)$;
and (v) a rough but simple analytic estimate of the average event duration $\langle t \rangle_{b}$,
which  generally is sensitive to details of the potential.
All of the analytic results are confirmed by transition-path-sampling simulations.

A number of interesting questions remain open. Perhaps most centrally, what changes in 
$\rho_b$ can be expected for high-dimensional system with potentially ``rough'' and
complex landscapes. A particular interest is in conformational 
transitions in proteins
~\cite{Zuckerman-Calmodulin_Go-JPCB-2004}.
Furthermore, it is relevant to consider how non-white noise affect event durations. 
Finally, how can the detailed relationship between $\rho_b$
and first passage-time be quantified, if at all?

\section*{ACKNOWLEDGMENTS}
We are grateful to Profs.\ Marty Ytreberg,
Joseph Rudnick, Daniel Boyanovsky, Vladimir Savinov and Mark Dykman for their interest 
and helpful comments and communications.
It also a pleasure to thank Drs.\ Arun Setty, Ed Lyman, and Svetlana Aroutiounian 
for helpful discussions. This work was supported in part 
by the NIH Grant GM070987 to DMZ.

\newpage
\bibliographystyle{h-physrev3}
\bibliography{paper}

%
\newpage

%

\end{document}